\documentclass[twocolumn,english]{revtex4-1}
\pdfoutput=1
\usepackage[T1]{fontenc}
\usepackage[latin9]{inputenc}
\usepackage{bm}
\usepackage{amsmath}
\usepackage{amssymb}
\usepackage{graphicx}
\usepackage{babel}
\usepackage{color}

\begin{document}
\title{Hybrid  plasmon-magnon  polaritons in graphene-antiferromagnet heterostructures} 

\author{Y. V. Bludov$^1$,  J. N. Gomes$^1$,   G. A. Farias$^3$, J. Fern\'andez-Rossier$^{2,4}$, M. I. Vasilevskiy$^{1,2}$,
\footnote{On leave from Departamento de F\'{i}sica Aplicada, Universidad de Alicante, Spain}
N. M. R. Peres$^{1,2}$}
\affiliation{$^1$Department of Physics, Center of Physics, and QuantaLab, University
of Minho, Campus of Gualtar, 4710-057, Braga, Portugal}
\affiliation{$^2$QuantaLab,International Iberian Nanotechnology Laboratory (INL), Av. Mestre
Jos\'e Veiga, 4715-330 Braga, Portugal}
\affiliation{$^3$Departamento de F\'{i}sica, Universidade Federal do Cear\'a, Caixa Postal
6030, Campus do Pici, 60455-900 Fortaleza, Cear\'a, Brazil}
\affiliation{$^4$Departamento de F\'{i}sica Aplicada, Universidad de Alicante, Carretera de San Vicente del Raspeig 
03690 San Vicente del Raspeig,
Alicante, Espa\~na}

\begin{abstract}
We consider  a hybrid structure formed by graphene and an insulating antiferromagnet, separated by a dielectric of thickness up to  $d\simeq 500 \,nm$.
When uncoupled, both graphene and the antiferromagneic surface  host  their own polariton modes coupling the electromagnetic field with plasmons in the case of graphene, and with magnons in the case of the antiferromagnet. We show that the  hybrid structure can host two new types of  hybrid polariton modes.  First, a surface magnon-plasmon polariton  whose dispersion is radically changed by the carrier density of the graphene layer, including a change of sign in the group velocity. Second, a surface plasmon-magnon polariton formed as a linear superposition of graphene surface plasmon and the antiferromagnetic bare magnon. This polariton has a dispersion with two branches, formed by the anticrossing between the dispersive surface plasmon and the magnon.   We discuss the potential these new modes have for combining photons, magnons, and  plasmons to reach new functionalities.
\end{abstract}
\maketitle

\section{Introduction}

Plasmons, excitons, phonons, and magnons are typical examples of collective excitations in condensed matter systems.  They all imply the presence of  poles with  frequencies $\Omega_p$  in the spectrum of the
response function  that describes the interaction of the system with electromagnetic waves.  As a result, the 
 propagation of electromagnetic waves with frequency $\omega$  in a material that hosts these collective modes is strongly modified, or even suppressed altogether,    for $\omega \simeq  \Omega_p$.  This general physical phenomenon  is rationalized in terms of the formation of new collective modes known as polaritons. 
 
 Quantum mechanically, polaritons are described as hybrid collective excitations that are linear superpositions of  a matter collective excitation and a photon.  Semiclassically,  they are described using Maxwell equations and constitutive relations that include the frequency dependent response functions.  In both instances, the underlying physical phenomenon is the emergence of a new type of wave or excitation, with properties different from those of the constituent collective mode and electromagnetic wave. 
 
Excitons, phonons, and plasmons, couple  predominantly to the electric component of the  electromagnetic  wave. In contrast, for magnon-polaritons,  it is the magnetic field  that couples to the   spins.
 Most  of the work so far has focused on  polaritons  that couple electromagnetic waves with just one type of collective mode (excitons, phonons, plasmons, spin waves).   
 Interestingly,  the same electromagnetic field would couple both to the spin and charge sector in a system that hosts both spin and charge collective modes. The electromagnetic field  of polaritons would thus
  provide a coupling channel between excitations that are normally un-coupled. 
 
The fabrication of nanostructures offers a new arena to explore hybrid systems with collective modes in the spin and charge sectors, that could result in a new type of polariton,  mixing spin and charge collective modes.   Here we explore this possibility in a  system that seems easy to fabricate with state of the art techniques.  We consider the coupling of surface magnon polaritons of an uniaxial antiferromagnet (AF) to surface plasmon polaritons (SPPs) in graphene.

The antiferromagnetic resonance (AFMR)  frequency  in 
insulating unixaxial antiferromagnets, that ultimately determines the magnon-polariton frequency, occurs 
in the {\bf THz} range,  well above the typical GHz range for ferromagnetic resonance and, importantly,  within the spectral range of graphene SPPs. The difference between AFMR and ferromagnetic resonance arises from the fact that the former is determined by the interplay of exchange and anisotropy \cite{antiferro-Keffer1952-pr}, whereas the latter is only given by magnetic anisotropy, which is much smaller in most cases.  It has been shown \cite{smp-Camley1982-prb,smp-exp-Jensen1995-prl} that 
uniaxial AFs, such as FeF$_2$, host both bulk and surface magnon polaritons (SMPs). These surface polaritons decay exponentially as we move away from the antiferromagnet-dielectric interface.

\begin{figure}
\includegraphics[width=8.5cm]{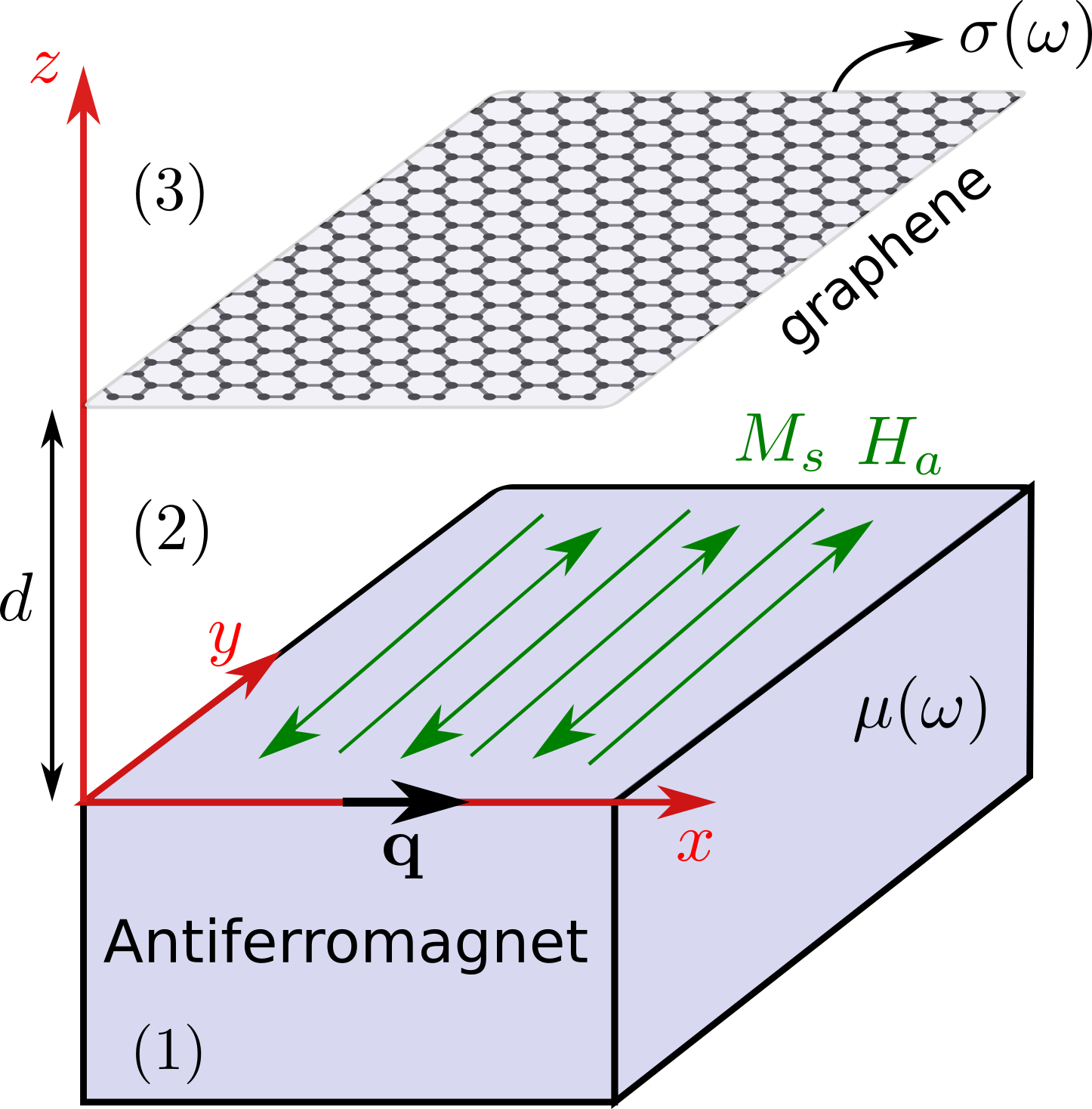}
\caption{Schematic drawing of the system considered in this work: a graphene sheet is located at 
a distance $d$ from the surface of an antiferromagnet, characterized by a magnetic permeability 
tensor ${\bm \mu}(\omega)$.
\label{fig:scheme}}
\end{figure}

The formation of hybrid modes occurs when the uncoupled modes are degenerate. Therefore, 
the existence of an experimental knob to tune the frequency of the modes is very convenient.  In the case of graphene,  gating  controls the carrier density,  leading to a  change of the  dispersion curve of SPPs. 
Therefore,  here we consider a graphene sheet at
a distance $d$ from the surface of an insulating AF, as shown in Fig. \ref{fig:scheme}.
Since both graphene and the insulating AF host their own polariton modes, here we 
explore 
whether this hybrid artificial material system hosts a new type of hybrid polariton that couples graphene surface plasmons and AF magnons at the same time.

In this work we show that indeed the tunability of the electromagnetic properties of an antiferromagnetic insulator can be achieved by gating a graphene
sheet (see Fig.\,\ref{fig:scheme}). In particular, we find a smooth transition from  the conventional regime where
the system has the energy propagation oriented along the same direction
of the SMP's wave vector to a regime where the energy flux is opposite to the wavevector, i.e. the group velocity of the hybrid excitation is negative.
If the dielectric layer between graphene and the antiferromagnet
has a negative electric permittivity, as it happens in a polar crystal near optical phonon resonances, a metamaterial \cite{metamaterial-Padilla2006-mt} with both negative
$\epsilon$ and $\mu$ can be achieved, thus exhibiting negative refraction.
Such a tunable system allows to control the direction of energy flow
at the surface of the antiferromagnet, thus providing a mechanism for directional propagation of the electromagnetic energy, without the need of an external magnetic field.

\section{Problem statement and main equations}
\label{sec:prob-stat}

\subsection{Antiferromagnetic  permeability}

The main objective of this paper is to investigate how the presence of graphene in the vicinity of an antiferromagnet influences the spectrum of SMPs, and {\it vice versa}, how the SPPs in graphene are affected by the antiferromagnet. Thus, we consider the semi-infinite AF, occupying the half-space $z<0$. The other half-space $z>0$ is supposed to be occupied by the vacuum with the graphene monolayer, arranged at plane $z=d$ parallel to the AF surface (see Fig.\,\ref{fig:scheme}).

The semi-infinite uniaxial antiferromagnet, such as FeF$_2$ or MnF$_2$\cite{smp-Camley1982-prb,smp-exp-Jensen1995-prl}, is described by the permeablilty tensor 
\begin{equation}
	\hat{\boldsymbol{\mu}}(\omega)=\left[\begin{array}{ccc}
	\mu_{a} & 0 &-i\mu_{b} \\
	0 & 1 & 0\\
	i\mu_{b} & 0 & \mu_a
	\end{array}\right].\label{eq:tensor-mu}
\end{equation}
The off-diagonal component
$\mu_{b}$ is finite in the presence of an external magnetic field  $H_{0}$, that
permits to tune the antiferromagnetic resonance frequency. In addition, application of  the latter provides
tunability of the resonance. In this work we only consider the case $H_0=0$, so that $\mu_b=0$ and
\cite{antiferro-Keffer1952-pr}:
\begin{equation}
\mu_{a}(\omega)=1+\frac{2\Omega_{s}^{2}}{\Omega_{0}^{2}-\omega^{2}},\label{eq:mu-isotropic}
\end{equation}

Here,   $\Omega_{0}=\gamma\mu_{0}\sqrt{2H_{a}H_{e}+H_{a}^{2}}$ is the antiferromagnetic resonance frequency,
that, unlike the case of uniaxial ferromagnets, depends not only on the anisotropy field  $H_a$, but as well on the exchange field $H_e$, which makes $\Omega_0$  much larger than the usual ferromagnetic resonance frequencies.

The gyromagnetic ratio is given by  $\gamma=e/(2m)$ where  $e$ and $m$ are the charge and mass of free electron, correspondingly.  The so called  saturation frequency is given by  $\Omega_{s}=\gamma\mu_{0}\sqrt{2H_{a}M_s}$,
where  $M_s$ denotes the saturation magnetization of each sublattice.

A calculation of the permeability tensor
for this system was performed long ago \cite{antiferro-Keffer1952-pr,antiferro-Kaganov1958-jetp,antiferro-Kaganov1962-jetp,smp-Mills1974-rpp,antiferro-Almeida1993-prb}. Equation  (\ref{eq:mu-isotropic}) can be obtained from a microscopic model Hamiltonian for 
spins, using both the spin wave approximation and Kubo formula for linear response to a transverse $ac$ field of frequency $\omega$ and zero wavevector.  Expression (\ref{eq:mu-isotropic}) is real, ignoring thereby losses. These could be included by replacing $\omega$ by $\omega+i\Gamma$, where $\Gamma$ describes a scattering rate.   

The spectral range for which $\mu_a(\omega)<0$ plays a very special role, as it becomes evident below. 
The condition $\mu_a(\omega)<0$ is met for $\Omega_0 <\omega<\sqrt{\Omega_0^2+2\Omega_s^2} $.

\subsection{Maxwell equations and boundary conditions}

The electromagnetic waves in such a layered structure are governed by macroscopic Maxwell equations, 

\begin{align}
&\mathrm{rot}\boldsymbol{\mathcal{H}}=
\frac{\partial\boldsymbol{\mathcal{D}}}{\partial t}+\boldsymbol{\mathcal{J}}^{(2D)}\delta\left( z-d\right) \label{eq:Maxwell-1}\\
&\mathrm{rot}\boldsymbol{\mathcal{E}}=-\frac{\partial\boldsymbol{\mathcal{B}}}{\partial t}\\
&\mathrm{div}\boldsymbol{\mathcal{D}}=\rho^{(2D)}\delta\left( z-d\right)\label{eq:Maxwell-3}\\
&\mathrm{div}\boldsymbol{\mathcal{B}}=0,\label{eq:Maxwell-4}
\end{align}
where delta-functions in Eqs.\eqref{eq:Maxwell-1} and  \eqref{eq:Maxwell-3} describe the two-dimensional nature of charges $\rho^{(2D)}$   and current $\boldsymbol{\mathcal{J}}^{(2D)}$ in the graphene monolayer.  

Maxwell equations \eqref{eq:Maxwell-1}--\eqref{eq:Maxwell-4}
can be solved separately in three spatial domains $z<0$, $0<z<d$ and $z>d$, which further in the paper will be denoted by $j=1, \, 2$ and 3, correspondingly. In the framework of this formalism, the Maxwell equations have the form:  

\begin{align}
&\mathrm{rot}\boldsymbol{\mathcal{H}}^{(j)}=
\frac{\partial\boldsymbol{\mathcal{D}}^{(j)}}{\partial t} \label{eq:Maxwell-j-1},\\
&\mathrm{rot}\boldsymbol{\mathcal{E}}^{(j)}=-\frac{\partial\boldsymbol{\mathcal{B}}^{(j)}}{\partial t},\\
&\mathrm{div}\boldsymbol{\mathcal{D}}^{(j)}=0,\\
&\mathrm{div}\boldsymbol{\mathcal{B}}^{(j)}=0,\label{eq:Maxwell-j-4}
\end{align}
and media indices are added as superscripts to the electric and magnetic fields $\boldsymbol{\mathcal{E}}^{(j)}$, $\boldsymbol{\mathcal{H}}^{(j)}$	$\boldsymbol{\mathcal{D}}^{(j)}$, $\boldsymbol{\mathcal{B}}^{(j)}$. It is notable that charges and currents induced in graphene do not enter explicitly Eqs.\,\eqref{eq:Maxwell-j-1}. These quantities are present in boundary conditions, which couples the electromagnetic fields in media $j=2$ and $j=3$. The boundary conditions at graphene plane take the explicit form:
\begin{align}
&\left.\mathbf{u}_{z}\times\left(\boldsymbol{\mathcal{E}}^{(3)}-\boldsymbol{\mathcal{E}}^{(2)}\right)\right|_{z=d}  =0,\label{eq:bound-E-32}\\
&\left.\mathbf{u}_{z}\times\left(\boldsymbol{\mathcal{H}}^{(3)}-\boldsymbol{\mathcal{H}}^{(2)}\right)\right|_{z=d} =\boldsymbol{\mathcal{J}}^{(2D)},\label{eq:bound-H-32}\\
&\left.\left(\boldsymbol{\mathcal{D}}^{(3)}-\boldsymbol{\mathcal{D}}^{(2)}\right)\right|_{z=d}\cdot\mathbf{u}_{z} =\rho^{(2D)},\\
&\left. \left(\boldsymbol{\mathcal{B}}^{(3)}-\boldsymbol{\mathcal{B}}^{(2)}\right)\right| _{z=d}\cdot\mathbf{u}_{z} =0.
\end{align}
Here $\mathbf{u}_{z}$ is a unit vector in the direction $z$, ''$\times$'' and ''$\cdot$'' mean vector and scalar products, respectively. The antiferromagnet is insulating, and therefore has no free charges and currents. In addition, we are assuming there is no surface magnetization. As a result, the boundary conditions between media $j=1$  and $j=2$, at the surface of the antiferromagnet, can be written as:
\begin{align}
&\left.\mathbf{u}_{z}\times\left(\boldsymbol{\mathcal{E}}^{(2)}-\boldsymbol{\mathcal{E}}^{(1)}\right)\right|_{z=0}  =0,\label{eq:bound-E-21}\\
&\left.\mathbf{u}_{z}\times\left(\boldsymbol{\mathcal{H}}^{(2)}-\boldsymbol{\mathcal{H}}^{(1)}\right)\right|_{z=0} =0,\label{eq:bound-H-21}\\
&\left.\left(\boldsymbol{\mathcal{D}}^{(2)}-\boldsymbol{\mathcal{D}}^{(1)}\right)\right|_{z=0}\cdot\mathbf{u}_{z} =0,\\
&\left.
\left(\boldsymbol{\mathcal{B}}^{(2)}-\boldsymbol{\mathcal{B}}^{(1)}\right)\right| _{z=0}\cdot\mathbf{u}_{z} =0.
\end{align}

In the following we look for the equations describing  electromagnetic waves 
with the propagation vector $\mathbf{k}$ lying  in-plane. There are two cases, $\mathbf{k}$
parallel and perpendicular to the AF's staggered magnetization $M_s$, that we take along $y$ (see Fig. 1).
\subsection{In plane propagation perpendicular to the staggered magnetization}

We consider first   the case where  the electromagnetic wave propagates in the direction $x$,  perpendicular to the direction of magnetization. This means that the problem under consideration is uniform in the direction $y$ (i.e. $\partial/\partial y \equiv 0$), and Maxwell equations \eqref{eq:Maxwell-j-1}-\eqref{eq:Maxwell-j-4} can be decomposed into two independent subsystems, which correspond to TE and TM polarizations.
%
  The TE-polarized wave includes the $y$-component of the electric field $\boldsymbol{\mathcal{E}}^{(j)}$ as well as $x$- and $z$-components of the magnetic field $\boldsymbol{\mathcal{H}}^{(j)}$, i.e. 

\begin{align}
&\boldsymbol{\mathcal{E}}^{(j)}(x,z,t) =\mathbf{u}_{y}{\mathcal{E}}^{(j)}_{y}(x,z,t),\\
&\boldsymbol{\mathcal{H}}^{(j)}(x,z,t) =\mathbf{u}_{x}{\mathcal{H}}^{(j)}_{x}(x,z,t)+\mathbf{u}_{z}{\mathcal{H}}^{(j)}_{z}(x,z,t).
\end{align}
Here $\mathbf{u}_{x}$ and $\mathbf{u}_{y}$ are unit vectors in directions $x$ and $y$, respectively.
The second subsystem, describing the TM-polarized wave, possesses  $x$- and $z$-components of the electric field and $y$-component of the magnetic field, 
\begin{align}
&\boldsymbol{\mathcal{H}}^{(j)}(x,z,t) =\mathbf{u}_{y}{\mathcal{H}}_{y}^{(j)}(x,z,t),\\
&\boldsymbol{\mathcal{E}}^{(j)}(x,z,t) =\mathbf{u}_{x}{\mathcal{E}}_{x}^{(j)}(x,z,t)+\mathbf{u}_{z}{\mathcal{E}}_{z}^{(j)}(x,z,t).
\end{align}
Moreover, one can assume the temporal and spatial dependencies of the electromagnetic fields as those of a plane wave with frequency $\omega$, travelling along the $x$-axis with wave-number $k$, that is, we can write $\boldsymbol{\mathcal{E}}^{(j)}(x,z,t)=\mathbf{E}^{(j)}(z)\exp(ikx - i\omega t)$, $\boldsymbol{\mathcal{H}}^{(j)}(x,z,t)=\mathbf{H}^{(j)}(z)\exp(ikx - i\omega t)$. In this formalism the wave amplitudes $\mathbf{E}^{(j)}$ and $\mathbf{H}^{(j)}$ depend upon $z$-coordinate only. 

We now take  into account the constitutive relations:
\begin{align} 
&\boldsymbol{\mathcal{D}}^{(j)}=\varepsilon_{0}\mathbf{E}^{(j)}\exp(ikx - i\omega t), \label{eq:constitutive-D}\\ &\boldsymbol{\mathcal{B}}^{(1)}=\mu_{0}\hat{\boldsymbol{\mu}}\mathbf{H}^{(1)}\exp(ikx - i\omega t), \label{eq:constitutive-B1}\\ 
&\boldsymbol{\mathcal{B}}^{(j\ne 1)}=\mu_{0}\mathbf{H}^{(j\ne 1)}\exp(ikx - i\omega t).\label{eq:constitutive-B23}
\end{align}
Such form of the constitutive relations describes the fact that the dielectric permittivities of all three media are equal to vacuum permittivity $\varepsilon_{0}$, and the magnetic permeability tensor of antiferromagnetic medium ($j=1$) is equal to $\mu_{0}\hat{\boldsymbol{\mu}}$.

Under all these assumptions, Maxwell equations \eqref{eq:Maxwell-j-1}
for the TE-polarization take the form 
\begin{eqnarray}
&\frac{dH_{x}^{(j)}}{dz}-ikH_{z}^{(j)}=-i\omega\varepsilon_{0}E_{y}^{(j)},\label{eq:maxwell-amp-s-1}\\
&\frac{dE_{y}^{(j)}}{dz}
=-i\omega\mu_{0}\left[\mu_{a}\left(\omega\right)\delta_{j,1}+\left(1-\delta_{j,1}\right)\right]H_{x}^{(j)},\label{eq:maxwell-amp-s-2}\\
&ikE_{y}^{(j)}
=i\omega\mu_{0}\left[\mu_{a}\left(\omega\right)\delta_{j,1}+\left(1-\delta_{j,1}\right)\right]H_{z}^{(j)}\label{eq:maxwell-amp-s-3}
\end{eqnarray} 
where $\delta_{j,1}$ is the Kronecker delta.
Correspondingly, the Maxwell equations for the TM-polarization read:
\begin{align}
&\frac{d E_{x}^{(j)}}{d z}-ikE_{z}^{(j)}=i\omega\mu_{0}H_{y}^{(j)},\label{eq:maxwell-amp-px-1}\\
&\frac{d H_{y}^{(j)}}{d z}=i\omega\varepsilon_{0}E_{x}^{(j)},\label{eq:maxwell-amp-px-2}\\
&ikH_{y}^{(j)}=-i\omega\varepsilon_{0}E_{z}^{(j)},\label{eq:maxwell-amp-px-3}
\end{align}

It is crucial that Eqs.\,\eqref{eq:maxwell-amp-s-1}--\eqref{eq:maxwell-amp-s-3} for the TE-polarization involve only the  $\mu_{xx}=\mu_{zz}$ components of the magnetic permeability tensor $\hat{\boldsymbol{\mu}}(\omega)$ [see Eq.\,\eqref{eq:tensor-mu}]. 
As a consequence, the magnetic medium is effectively isotropic with respect to the TE-polarized wave, when electromagnetic wave propagates along $x$-direction (perpendicular to the staggered magnetization). At the same time, only $yy$ component of the magnetic permeability tensor $\hat{\boldsymbol{\mu}}(\omega)$ is present in the Maxwell equations for the TM-polarized wave \eqref{eq:maxwell-amp-px-1}, which is equal to unity [see Eq.\,\eqref{eq:tensor-mu}]. Therefore, we would expect that the AF medium in the structure depicted in Fig.\,\ref{fig:scheme} would not exert any influence on the spectrum of the TM-polarized wave.
As we will see this is not exactly the case near the  resonance frequency $\Omega_0$.

\subsection{In plane propagation parallel  to the staggered magnetization}

We now consider the propagation along the $y$ direction, parallel to the staggered magnetization.
 In this case the homogeneity of the system under consideration in the direction $x$ ($\partial/\partial x\equiv 0$) also implies the separation of Maxwell equations \eqref{eq:Maxwell-j-1}-\eqref{eq:Maxwell-j-4} into the TE subsystem:

\begin{align}
&ikH_{z}^{(j)}-\frac{dH_{y}^{(j)}}{dz}=-i\omega\varepsilon_{0}E_{x}^{(j)},\label{eq:maxwell-amp-sy-1}\\
&\frac{dE_{x}^{(j)}}{dz}
=i\omega\mu_{0}H_{y}^{(j)},\label{eq:maxwell-amp-sy-2}\\
&-ikE_{x}^{(j)}
=i\omega\mu_{0}\left[\mu_{a}\left(\omega\right)\delta_{j,1}+\left(1-\delta_{j,1}\right)\right]H_{z}^{(j)}\label{eq:maxwell-amp-sy-3}
\end{align}

 and the TM subsystem:  

\begin{align}
&ikE_{z}^{(j)}-\frac{d E_{y}^{(j)}}{d z}=\label{eq:maxwell-amp-p-1}\\
&=i\omega\mu_{0}\left[\mu_{a}\left(\omega\right)\delta_{j,1}+\left(1-\delta_{j,1}\right)\right]H_{x}^{(j)},\\
&\frac{d H_{x}^{(j)}}{d z}=-i\omega\varepsilon_{0}E_{y}^{(j)},\label{eq:maxwell-amp-p-2}\\
&-ikH_{x}^{(j)}=-i\omega\varepsilon_{0}E_{z}^{(j)}\,.
\label{eq:maxwell-amp-p-3}
\end{align}
 
While obtaining these equation, we used the plane-wave spatio-temporal dependence of the field $\boldsymbol{\mathcal{E}}^{(j)}(y,z,t)=\mathbf{E}^{(j)}(z)\exp(iky - i\omega t)$, $\boldsymbol{\mathcal{H}}^{(j)}(y,z,t)=\mathbf{H}^{(j)}(z)\exp(iky - i\omega t)$, as well as the constitutive relations, similar to Eqs.\,\eqref{eq:constitutive-D} (except the dependence upon $y$-coordinate instead of $x$). As a result, the antiferromagnet, whose response involves components $yy$ and $zz$ of the magnetic permeability tensor \eqref{eq:tensor-mu}, is effectively anisotropic with respect to the TE-polarized waves [see Eqs.\,\eqref{eq:maxwell-amp-sy-1}]. Furthermore, the AF medium influences the properties of the TM-polarized waves [see Eq.\,\eqref{eq:maxwell-amp-p-1}].      

\section{Uncoupled modes:  surface plasmon-polaritons and surface magnon-polaritons}

In this section we briefly revisit the properties of the SPPs in graphene, on one side, and the SMPs in the AF in the other, ignoring their mutual coupling. This provides a background to understand the nature of the new hybrid collective modes that arise in the combined graphene/AF structure. 
We keep the discussion at a qualitative level. The quantitative theory presented in the next sections includes, as limiting cases, a theoretical description of these excitations. 

\begin{table}
	\hspace{-0.5cm}
	\vline
	\begin{tabular}{c|c|c|c|}
		\hline 
		{\bf Surface polariton type}  &{\bf System}           &{\bf Pol.} & ${\bf Wavevector}$ \\ 
		\hline
		Magnon  & AF   & TE & $\mathbf{k}\cdot\mathbf{M}_{s}=0$ \\
		\hline
		Plasmon & G  &  TM  & isotropic \\
		\hline 
		{Magnon-plasmon} & AF+G  & TE & $\mathbf{k}\cdot\mathbf{M}_{s}=0$  \\
		\hline
		{Plasmon-magnon} & AF+G  & TM & $\mathbf{k}\times\mathbf{M}_{s}= 0$  \\
		\hline		
	\end{tabular}
	\caption{Summary of the different surface polariton excitations discussed in this work. The type of the polariton indicates the elementary excitation coupled to the EM field, except for the third line where plasmons are not directly involved in the hybrid wave.}
	\label{tab:parameters}
\end{table}

\subsection{Magnon--polaritons}
The case of bulk magnon-polaritons  for a uniaxial antiferromagnet was studied by \cite{smp-Camley1982-prb}. It was found that only TE modes exist, with a dispersion relation that we derive in  the appendix \ref{app} 
and is depicted in Fig.\ref{fig:Schematic-dispersion-relation} by dashed red lines.  The magnon--polariton dispersion is mathematically identical to the case of Hopfield exciton--polaritons in a semiconductor. Magnon--polaritons come in two branches [acoustical $\omega_a(k)$ and optical $\omega_o(k)$], both twice degenerate on account of the dimension of the symmetry plane perpendicular to the easy axis.   At frequencies far from the AFMR resonance, $\omega\lessgtr\Omega_0$ these two branches are close to the photon dispersion curve $\omega= ck$, while in the frequency range $\omega\lesssim\Omega_0$ the lowest, acoustic branch asymptotically approaches the AFRM frequency as $k\to\infty$, i.e. $\omega_a(\infty)=\Omega_0$. In the vicinity of the AFMR frequency two modes are separated by the frequency gap, whose value is roughly given by $\omega_o(0)-\omega_a(\infty)\approx\Omega_s^2/\Omega_0$.

\begin{figure}[hbt]
\includegraphics[width=8.5cm]{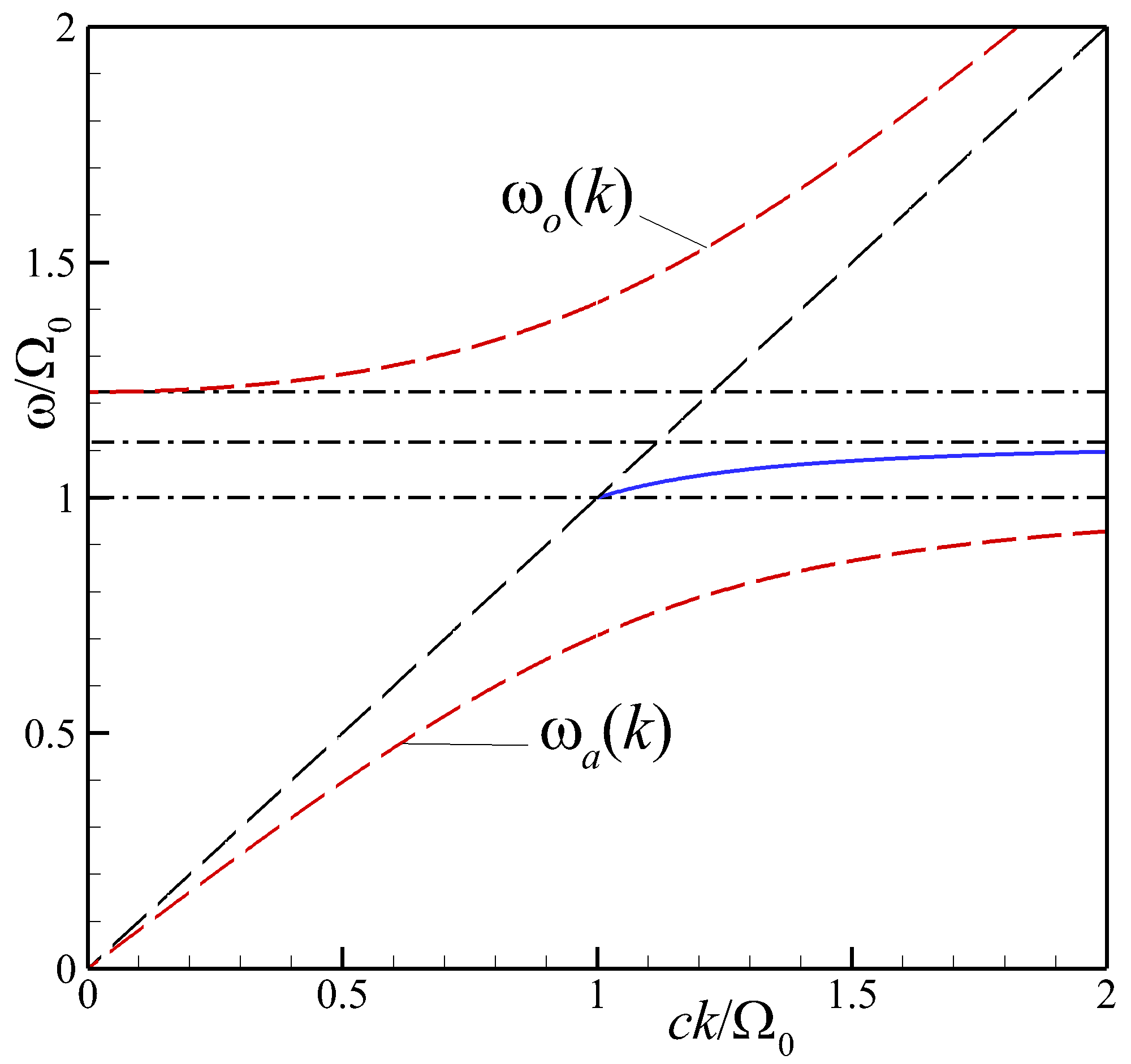}
\caption{Schematic dispersion relation of surface (blue solid line) and bulk (red dashed lines) magnon-polaritons in the system without graphene, $E_F=0$, and with the AF parameters $\Omega_{s}=0.5 \Omega_{0}$.  The black dashed line corresponds to the vacuum light line $\omega=kc$, while frequencies $\Omega_0$, $\sqrt{\Omega_0^2+\Omega_s^2}$, and $\sqrt{\Omega_0^2+2\Omega_s^2}$ are depicted by horizontal dash-and-dotted black lines (from bottom to top, respectively).
	\label{fig:Schematic-dispersion-relation}}
\end{figure}

At the surface of AF, the collective excitations of spins, i.e. magnons can be coupled to an electomagnetic wave, forming surface magnon-polaritons  (SMP). The key property of SMPs [see the first line of Table \ref{tab:parameters}] is that they are TE-polarized waves and it  was first considered
by Camley e Mills \cite{smp-Camley1982-prb} (see also Ref. [\onlinecite{smp-ni-Tarkhanyan2009-pssb}]). One of  the first reported \cite{smp-exp-Jensen1995-prl}
observations of SMPs was in the antiferromagnet FeF$_2$ using the technique of attenuated total reflection. The same method first used to observe SPPs in metallic-dielectric interfaces.
These surface magnon--polaritons [depicted by blue solid line in Fig.\,\ref{fig:Schematic-dispersion-relation}] only exist for $ck >\Omega_0$, as we show below [see subsection \ref{subsec:smp}].

In Sec.\,\ref{sec:smp} we study how the interaction between electromagnetic field of SMP at vacuum/AF interface and {\it forced} charge-carrier oscillations in graphene modify the SMP spectrum. The resulting hybrid mode will be referred to as surface magnon-plasmon polariton and its fundamental properties are summarized in the third line of Table \ref{tab:parameters}.

\subsection{Graphene surface plasmon-polaritons}
Graphene SPPs can be understood as solutions of the Maxwell equations that describe an electromagnetic wave propagating along a conductive graphene sheet. The electromagnetic field is strongly confined in the neighbourhood of the graphene, with evanescent off-plane tails.  Graphene SPPs are TM-polarized and their dispersion curve, $\omega(k)<ck$, can be tuned by controlling the carrier density. Fundamental properties of SPPs in graphene are briefly summarized in the second line of Table \ref{tab:parameters}.

Compared to the SPPs at the surface of noble metals, the graphene polaritons have  significant advantages since they are characterized by both a longer lifetime and a higher degree of field confinement \cite{spp-gr-rev-Koppens2011-nl,spp-gr-Nikitin2011-prb}. If graphene layer is deposited on a polar substrate, the electromagnetic field of graphene SPPs can interact with optical phonons in the substrate, thus forming hybrid modes called surface plasmon-phonon-polaritons \cite{spp-gr-phon-Brar2014-nl,spp-gr-phon-Liu2010-prb,spp-gr-phon-Luxmoore2014-acsphot,spp-gr-phon-Fei2011-nl,kumar2015}. 
Here, we also expect hybrid polaritons invloving two physically distinct elementary excitations in the materials involved. This mode will be called surface plasmon-magnon polariton (SPMP) and its properties are summarized in the forth line of Table \ref{tab:parameters}. The study of this mode will be considered in detail in Sec.\,\ref{sec:mod-spp}. However, the system considered in the present work is different from surface plasmon-phonon-polaritons in one important aspect. The two materials combined in our system, if taken separately, support surface waves whose polarizations are orthogonal to each other.

From Table \ref{tab:parameters}, it is apparent that graphene SPPs are TM modes whereas   AF's  surface magnon--polaritons are TE modes. Therefore,  in order to study the polaritons of the hybrid system, we need to consider both TM and TE modes.

\section{Surface magnon-plasmon polaritons}
\label{sec:smp}

\subsection{General equations for TE modes}

 In this Section we shall demonstrate that interaction between magnons and free charges in graphene via electromagnetic radiation modifies the spectrum of SMP. 
Boundary condition on graphene (\ref{eq:bound-H-32}) couples the in-plane components of the electric and magnetic fields; in the case opf TE modes it involves transverse plasmons in graphene. For such plasmons, the current is perpendicular to the wavevector, the charge density is kept constant\cite {Stauber2013} and they do not interact with the electromagnetic radiation directly. However, in the hybrid structure considered here they can interact indirectly, through the AF whose magnons do couple to the TE-polarized radiation. Such hybrid evanescent waves will be called surface magnon--plasmon--polaritons (SMPP). They propagate along the direction perpendicular to the staggered magnetization in the antiferromagnet. 

It should be noticed that the frequency range of the antiferromagnetic resonance lies in the THz spectral region, where interband transitions in graphene play no role. Therefore, we consider the optical response of graphene described by a Drude formula without losses:\cite {Note2}
\begin{equation}
\sigma(\omega)=i\frac{2 e^2}{h}\frac{E_F}{\hbar\omega}\,,\label{eq:drude-cond}
\end{equation} 
with $E_F$ being the Fermi energy of doped graphene. This equation is valid as long as the Fermi energy is much larger than $k_BT$ ($k_B$ is the Boltzmann constant and $T$ the temperature).

The propagation of TE-polarized waves is governed by the Maxwell equations in the form \eqref{eq:maxwell-amp-s-1}--\eqref{eq:maxwell-amp-s-3}.
Substitution of Eqs.\,\eqref{eq:maxwell-amp-s-2} and \eqref{eq:maxwell-amp-s-3} into Eq.\,\eqref{eq:maxwell-amp-s-1} results into the Helmholtz equation,

\begin{align}
&-\frac{d^2E_{y}^{(j)}}{dz^2}+k^2E_{y}^{(j)}\nonumber\\
&=\frac{\omega^2}{c^2}\left[\mu_{a}\left(\omega\right)\delta_{j,1}+\left(1-\delta_{j,1}\right)\right]E_{y}^{(j)},\label{eq:helmholtz-s}
\end{align}
whose solution in semi-infinite media $j=1$ and $j=3$ can be expressed as

\begin{align}
&E_{y}^{(1)}\left(z\right)=E_{y}^{(1)}\left(0\right)\exp(\beta^{(1)}z),	\label{eq:Ey1}\\	
&E_{y}^{(3)}\left(z\right)=E_{y}^{(3)}\left(d\right)\exp\left[-\beta^{(2)}\left(z-d\right)\right],	\label{eq:Ey3}	
\end{align}
with $E_{y}^{(1)}\left(0\right)$ and $E_{y}^{(3)}\left(d\right)$ being the values of electric field at the surface of the antiferromagnet and graphene, respectively, and
\begin{align}
&\beta^{(1)}=\sqrt{k^{2}-\omega^{2}\mu_{a}\left(\omega\right)/c^{2}},\label{eq:beta-1}\\
&\beta^{(2)}=\sqrt{k^{2}-\omega^{2}/c^{2}}.\label{eq:beta-2}
\end{align}

In the considered framework we assume, for simplicity, that both the antiferromagnet and graphene are lossless media, which means that both the in-plane wavevector $k$ and frequency $\omega$ have real and positive values. Moreover, since we are interested in studying the surface wave, $\beta^{(1)}$ and $\beta^{(2)}$ are also real and characterize the  inverse penetration length of the evanescent fields.
The respective signs of the exponents in Eqs.\,\eqref{eq:Ey1} and \eqref{eq:Ey3} were chosen to satisfy the boundary conditions at $z=\pm\infty$ [namely, $E_{y}^{(1)}\left(-\infty\right)=E_{y}^{(3)}\left(\infty\right)=0$], which describe the absence of modes, growing exponentially towards $|z|\to\infty$.

The respective magnetic fields in media $j=1$ and $j=3$ can be obtained by substituting Eqs.\,\eqref{eq:Ey1} and  \eqref{eq:Ey3} into Eqs.\,\eqref{eq:maxwell-amp-s-2} and \eqref{eq:maxwell-amp-s-3}, and expressing the fields as

\begin{align}
&H_{x}^{(1)}\left(z\right)=-\frac{\beta^{(1)}}{i\omega\mu_0\mu_a\left(\omega\right)}E_{y}^{(1)}\left(0\right)
e^{z_1},\label{eq:Hx1}\\
&H_{z}^{(1)}\left(z\right)=\frac{k}{\omega\mu_0\mu_a\left(\omega\right)}E_{y}^{(1)}\left(0\right)
e^{z_1},\label{eq:Hz1}\\
&H_{x}^{(3)}\left(z\right)=\frac{\beta^{(2)}}{i\omega\mu_0}E_{y}^{(3)}\left(d\right)
e^{-(z_2-d_2)}
,\label{eq:Hx3}\\
&H_{z}^{(3)}\left(z\right)=\frac{k}{\omega\mu_0}E_{y}^{(3)}\left(d\right)
e^{-(z_2-d_2)},\label{eq:Hz3}
\end{align}
where we have adopted the notation $z_j\equiv \beta^{(j)} z$, $d_j\equiv \beta^{(j)} d$ with $j=1,2$.

For the spacer layer $j=2$ that separates  graphene and the  AF's surface, 
$0<z<d$, 
there are no restrictions on the presence of exponentially growing or decaying modes.
As a result, the solution is composed as the superposition of these two modes. In terms of hyperbolic functions and amplitudes of electric field at boundaries $z=0$ and $z=d$, this solution can be represented as 
\begin{align} 
&E_{y}^{(2)}(z) =E_{y}^{(2)}(0){\cal F}_S(d_2-z_2)
+E_{y}^{(2)}(d){\cal F}_S(z_2), \label{eq:Ey2} \\
%
%
&H_{x}^{(2)}(z)=\eta_x \left[E_{y}^{(2)}(0){\cal F}_C(d_2-z_2) \right.\\
&\left. -E_{y}^{(2)}(d) {\cal F}_C(z_2) \right],
\label{eq:Hx2}\\
&H_{z}^{(2)}(z)=\eta_z \left[E_{y}^{(2)}(0)
{\cal F}_S(d_2-z_2) +E_{y}^{(2)}(d) {\cal F}_S(z_2)\right],
\label{eq:Hz2}
\end{align}

where we have defined

\begin{align}
&{\cal F}_S(z_2)\equiv\frac{\sinh(z_2)}{\sinh(d_2)}, \,\,\,\;\;
{\cal F}_C(z_2)\equiv\frac{\cosh(z_2)}{\sinh(d_2)},  \\
&\eta_x\equiv \frac{\beta^{(2)}}{i\omega\mu_0}
, \,\,\,\;\;
\eta_z=\frac{k}{\omega\mu_0}.
\end{align}

For the particular case of TE-polarized wave (and plane wave temporal and spatial dependencies, mentioned above), first and second relations in the boundary conditions \eqref{eq:bound-E-32} and \eqref{eq:bound-H-21} are expressed as

\begin{align}
&E_{y}^{(3)}\left(d\right)=E_{y}^{(2)}\left(d\right),\label{eq:bound-E-32-s}\\
&H_{x}^{(3)}\left(d\right) - H_{x}^{(2)}\left(d\right)=\sigma(\omega)E_{y}^{(2)}\left(d\right),\label{eq:bound-H-32-s}\\
&E_{y}^{(2)}\left(0\right)=E_{y}^{(1)}\left(0\right),\label{eq:bound-E-21-s}\\
&H_{x}^{(2)}\left(0\right) - H_{x}^{(1)}\left(0\right)=0.\label{eq:bound-H-21-s}
\end{align}

It should be pointed out that boundary condition \eqref{eq:bound-H-32-s} was obtained by using the two-dimensional current in graphene $\boldsymbol{\mathcal{J}}^{(2D)}$ expressed via the graphene conductivity and electric field as 
\begin{eqnarray}
\boldsymbol{\mathcal{J}}^{(2D)}=\mathbf{u}_y\sigma\left(\omega\right)E_y^{(2)}\left(d\right)\exp\left(ikx-i\omega t\right).
\end{eqnarray}
Substitution of Eqs.\,\eqref{eq:Hx1} and \eqref{eq:Hx2} into boundary conditions \eqref{eq:bound-H-32-s} results into the homogeneous linear system of equations:
\begin{eqnarray}
\left(\begin{array}{cc}
 {\cal M}_{11}& 
 \frac{-\beta^{(2)}}{\sinh(d_2)} \\
 \frac{-\beta^{(2)}}{\sinh(d_2)} & 
{\cal M}_{22}
\end{array}\right)
\left(\begin{array}{c} E_{y}^{(2)}\left(d\right)
 \\
E_{y}^{(2)}(0)
\end{array}\right)=
\left(\begin{array}{c} 0
 \\
0
\end{array}\right),
\label{eqmat}
\end{eqnarray}
where 
\begin{eqnarray}
{\cal M}_{11}&&\equiv  \beta^{(2)}+\frac{\beta^{(2)}}{\tanh(d_2)}- i\omega\mu_0\sigma\left(\omega\right),
\\
{\cal M}_{22}&&\equiv \frac{\beta^{(2)}}{\tanh(d_2)}+\frac{\beta^{(1)}}{\mu_a\left(\omega\right)}.
\end{eqnarray}

Non trivial solutions of  Eq.\,(\ref{eqmat}) occur when 
the determinant of the matrix vanishes:
\begin{equation}
{\cal M}_{11}{\cal M}_{22}- \left(\frac{\beta^{(2)}}{\sinh(d_2)}\right)^2=0
\label{eq:dr-s}
\end{equation}
For a given value of $k$, this equation can be satisfied by several values of $\omega_{\nu}(k)$ that define
the different polariton modes in the system, which are distinguished by the index $\nu$.
In the following we discuss these polariton modes  in two cases.  First, we take the $d\to\infty$ limit, where there is no coupling
between graphene and the AF's surface.  This permits to recover an expression for the surface magnon-polariton \cite{smp-Camley1982-prb}.  Later we shall also take the limit where $\beta^{(2)}d$ is small, for which the presence of graphene
modifies the SMP properties.

\subsection{No dressing: the $d\to\infty$ limit}\label{subsec:smp}
In the case of infinite distance $d\to\infty$ between the antiferromagnet surface and graphene monolayer, the dispersion relation \eqref{eq:dr-s} transforms into 
\begin{eqnarray}
\left[ 2\beta^{(2)}-i\omega\mu_0\sigma\left(\omega\right)\right]\left[\beta^{(2)}+\frac{\beta^{(1)}}{\mu_a\left(\omega\right)}\right]=0. \label{eq:dr-s-decoupled}
\end{eqnarray}
The term in the first braces of Eq.\,\eqref{eq:dr-s-decoupled} is always positive owing to the positiveness of the imaginary part of Drude conductivity \eqref{eq:drude-cond}\footnote{If the interband transitions are taken into account in the expression of graphene's conductivity, its imaginary part can be negative. In this case graphene is able to sustain s-polarized surface plasmon-polaritons\cite{gr-spp-te-Mikhailov2007-prl}, which dispersion relation is described by the term in first braces of Eq.\,\eqref{eq:dr-s-decoupled}. Nevertheless, their typical frequencies lies in the vicinity of double graphene's Fermi energy $\omega\sim2E_F$, i.e. significantly higher than the antiferromagnet resonance frequency $\Omega_0$.},  while setting equal to zero the second term in braces in \eqref{eq:dr-s-decoupled} yields the dispersion relation of the SMP existing at single interface between vacuum and antiferromagnet\cite{smp-Mills1974-rpp}.

Since both $\beta^{(1)}$ and $\beta^{(2)}$, defined in Eqs.\eqref{eq:beta-1},  are positive, Eq.\eqref{eq:dr-s-decoupled} only has solutions when $\mu_{a}\left(\omega\right)<0$, i.e. in the aforementioned frequency range $\Omega_0\le\omega\le\sqrt{\Omega_0^2+2\Omega_s^2}$. 
 The simultaneous positiveness of the arguments of $\beta^{(1)}$ and $\beta^{(2)}$ in that range takes place when $k\ge\omega/c$, i.e., at the right of the "light line" $\omega=ck$ (which is depicted in Fig.\,\ref{fig:Schematic-dispersion-relation} by black dashed line). 
  SMP's dispersion relation \eqref{eq:dr-s-decoupled} after some algebra can be expressed in the explicit form
\begin{eqnarray}
\begin{aligned}
&\omega=\left[\Omega_{s}^{2}+\frac{\Omega_{0}^{2}}{2}+c^{2}k^{2}\right .\\
&\left . -\sqrt{\left(c^{2}k^{2}-\frac{\Omega_{0}^{2}}{2}\right)^{2}+\Omega_{s}^{2}\left(\Omega_{0}^{2}+\Omega_{s}^{2}\right)}\right]^{1/2},
\end{aligned}
\end{eqnarray}
which is depicted in Fig.\,\ref{fig:Schematic-dispersion-relation} by solid blue line. The SMP's spectrum starts on the light line at frequency $\omega=\Omega_0$ and wavevector $k=\Omega_0/c$.
In the vicinity of this point, the dispersion relation  is described  approximately by the relation 
\begin{eqnarray}
\omega\approx\Omega_{0}+\frac{2\Omega_{s}^{2}}{\Omega_{0}^{2}+2\Omega_{s}^{2}}\left(ck-\Omega_{0}\right).	
\end{eqnarray}

At large wavevectors, $k\to \infty$, the SMP's spectrum asymptotically approaches the frequency $\omega=\sqrt{\Omega_0^2+\Omega_s^2}$ as 
\begin{eqnarray}
\omega\approx\sqrt{\Omega_{s}^{2}+\Omega_{0}^{2}}\left[1-\frac{\Omega_{s}^{2}}{4c^{2}k^{2}}\right].
\end{eqnarray}

Expectedly, the SMP's spectrum appears in the gap between two branches of the TM-polarized bulk magnon-polariton dispersion relation, 
\begin{equation}
\beta^{(1)}=0,\label{eq:bulk-dr}
\end{equation}
which is depicted in Fig.\,\ref{fig:Schematic-dispersion-relation} by red dashed lines (see Appendix \ref{app}).

\subsection{Dressing: the  finite $\beta^{(2)}d<1$  case}

\begin{figure}
\includegraphics[width=8.5cm]{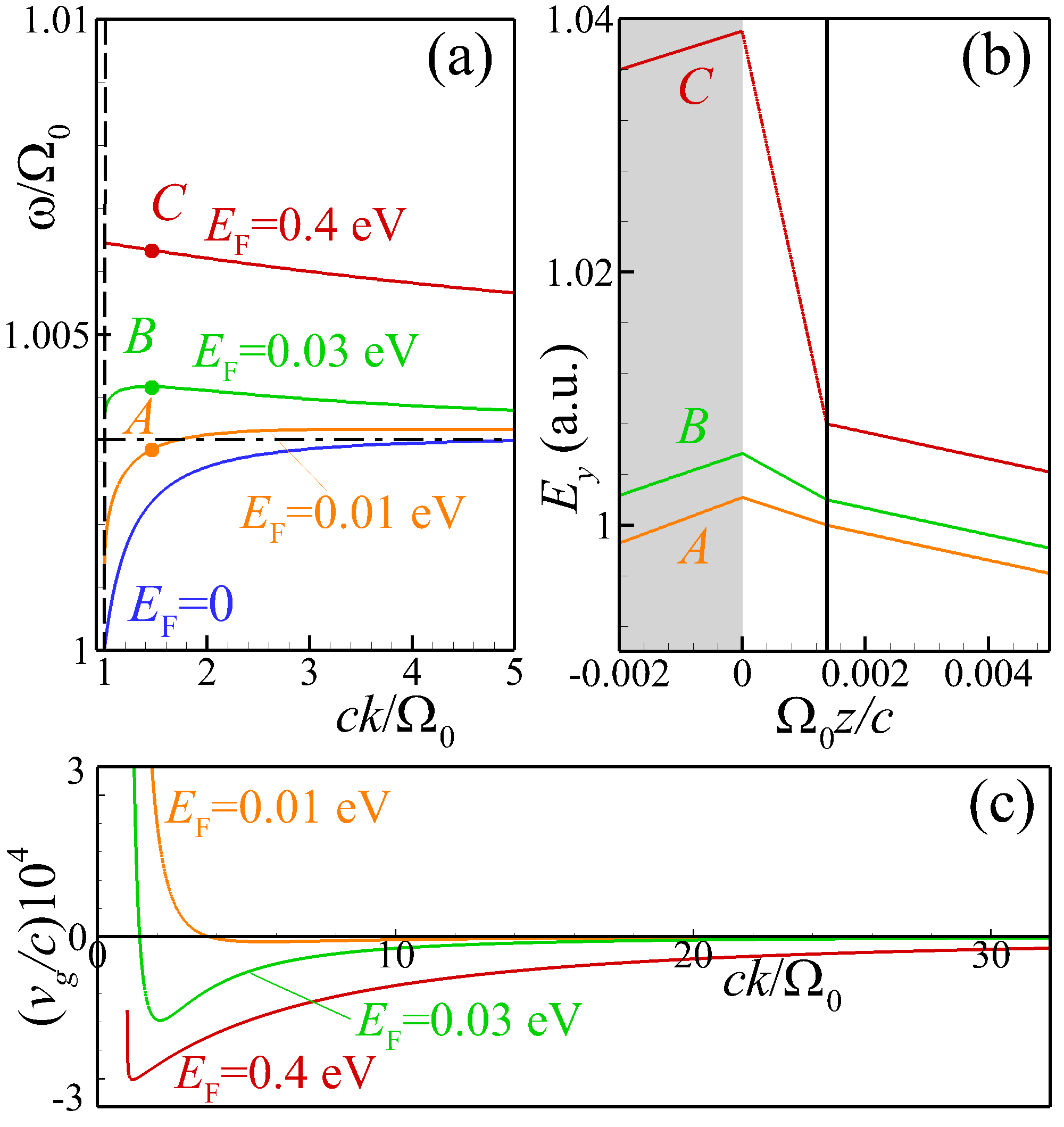}
\caption{(a) Surface magnon--plasmon--polariton (SMPP) spectrum in the AF/graphene structure for Fermi energy values $E_{F}=0$ (solid blue line), $0.01\,$eV (solid orange line), $0.03\,$eV (solid green line), and $0.4\,$eV (solid red line). Black dashed line stands for the light line in vacuum, $\omega=ck$; (b) Spatial distributions of the SMPP electric field for the modes with $ck/\Omega_0=1.46$ and $0.01\,$eV (orange line A), $0.03\,$eV (green line B), and $0.4\,$eV  (red line C). These modes are indicated in panel (a) by the respective letters A, B and C. The region occupied by the antiferromagnet is shadowed in panel (b) and the position of graphene is shown by vertical bold black solid line; (c) Group velocity, $v_g=d\omega/dk$ (in dimensionless units $v_g/c$) of the SMPP modes with $E_F=0.01\,$eV (orange line), $0.03\,$eV (green line), and $0.4\,$eV (red line). In all panels the fields and magnetization of the antiferromagnet are $\mu_0H_{a}=0.787\,{\rm T}$, $\mu_0H_{e}=55.3\,{\rm T}$, and $\mu_0M_{s}=0.756\:{\rm T}$, for the antiferromagnet MnF$_{2}$ [\onlinecite{antiferro-Macedo2017}]. The spacer between the antiferromagnet and graphene has thickness $d=500\,$nm. 
The magnitude of the fields was chosen arbitrarily for convenient visualization of their profiles.
\label{fig:Spectrum-of-the-smp-graphene}}
\end{figure}

We now discuss the influence of graphene on the properties of the SMP. This happens when
the distance between antiferromagnet and graphene is  such that  $\beta^{(2)}d$ is not very large and the Fermi energy in graphene is not at the Dirac point, $E_F\neq 0$. 
In this case the spectrum of SMPs is strongly modified owing to the influence of free charges in graphene on the electromagnetic field of the SMP, supported by the surface of the AF. 
The SMP spectrum \eqref{eq:dr-s} for relatively small distance $d=500\,$nm is depicted in Fig.\, \ref{fig:Spectrum-of-the-smp-graphene}(a) for different values of the Fermi energy. Thus, for finite doping of the graphene,  the dressed SMP spectrum has a starting-point with the frequency $\omega>\Omega_0$, lying on the light line. An increase of the Fermi energy $E_F$ results into the shift of the starting-point of the spectrum towards   higher frequencies. An expression for the starting-point frequency 
\begin{eqnarray}
\omega_i=\left[\frac{\Omega_{0}^{2}a-2b\left(2\Omega_{s}^{2}+\Omega_{0}^{2}\right)}{2\left(a-b\right) }\right.\nonumber\\
\left.+\frac{\sqrt{\Omega_{0}^{4}a^{2}+8ab\Omega_{s}^{2}\left(2\Omega_{s}^{2}+\Omega_{0}^{2}\right)}}{2\left( a-b\right) }\right]^{1/2}>\Omega_0,\label{eq:starting-point}
\end{eqnarray} 
can be obtained explicitly from the dispersion relation \eqref{eq:dr-s} by putting the condition $\beta^{(2)}=0$. In Eq.\,\eqref{eq:starting-point} $a=\left[1+4\alpha E_{F}d/(\hbar c)\right]^{2}$, $b=8\left[\alpha E_{F}/(\hbar\Omega_{s})\right]^{2}$, and $\alpha=e^2 / (4\pi\varepsilon_0 \hbar c)$ is the fine-structure constant.  It is apparent that results are independent of the sign of $E_F$, i.e., for graphene with extra electrons or holes.

In the $k\rightarrow \infty$ limit, the spectrum tends to $\omega= \sqrt{\Omega_0^2+\Omega_s^2}$. Therefore, as $E_F$ is ramped up  and the spectrum is pushed up in frequency at the smallest allowed values of $k$, so that the starting SMPP's frequency becomes larger than that limiting frequency ($\omega_i>\sqrt{\Omega_0^2+\Omega_s^2}$), their group velocity  $v_g=d\omega/dk$ has to be negative. 
This happens at  experimentally attainable dopings of the graphene. For  $E_F=0.01\,$eV and $E_F=0.03\,$eV, orange and green lines in Fig.\,\ref{fig:Spectrum-of-the-smp-graphene}(c), respectively, $v_g<0$ in a range of high values of $k$.  For higher values of $E_F$   [ $E_F=0.4\,$eV, red line in Fig.\,\ref{fig:Spectrum-of-the-smp-graphene}(c)] $v_g<0$ for all values of $k$.

This result is distinct from the zero Fermi energy case, where SMPP's group velocity [slope of the dispersion curve, $\omega(k)$, depicted by solid blue line in Fig.\,\ref{fig:Spectrum-of-the-smp-graphene}(a)] is positive in all range of frequencies and wavevectors. It is possible to see, that the group velocity is much smaller than the speed of light in vacuum, $c$. Even more, in short-wavelength limit $ck/\Omega_0\gtrsim 30$ the group velocity is less than 10$^{-5}c$, i.e. SMPPs are slow waves.

Examples of spatial profiles of SMPP modes are shown in Fig.\,\ref{fig:Spectrum-of-the-smp-graphene}(b). As can be seen from the figure, in the case of TE-polarized wave $E_y(0)>E_y(d)$, so the field is mainly concentrated nearby of the antiferromagnet surface. From the comparison of lines A, B, and C in Fig.\,\ref{fig:Spectrum-of-the-smp-graphene}(b) it is possible to conclude that higher graphene doping  level leads to a stronger localization of the electromagnetic field in the vicinity of the antiferromagnet surface.
 
\section{Surface plasmon-magnon polaritons}
\label{sec:mod-spp}
 \subsection{Equations for TM modes}

We now consider the case of TM-polarization, for which  the graphene layer is able to sustain SPPs, oscillations of charge-carrier density in graphene coupled to the electromagnetic radiation\cite{gr-spp-Jablan2009-prb,spp-gr-rev-Bludov2013-ijmfb}.  We anticipate our main finding:   in the AF-graphene coupled structure, the graphene SPP is hybridized with the AF magnon, resulting in a polariton spectrum with 2 branches, that reflects the emergence of a hybrid collective mode that combines graphene plasmons with AF magnons.
We shall call these hybrid excitations surface plasmon--magnon--polaritons (SPMPs).
  
We address the case where the electromagnetic field is 
TM polarized and wave propagates along the $y$-axis, i.e. parallel to the staggerred magnetization. In this situation the electromagnetic field is defined by the Maxwell equations in the form of Eqs.\,\eqref{eq:maxwell-amp-p-1}.
Their solutions in the AF region can be 
 expressed as

\begin{align}
&H_{x}^{(1)}\left(z\right)=-i\frac{\omega\varepsilon_0}{\beta^{(1)}}E_{y}^{(1)}\left(0\right)\exp(z_1),\label{eq:Hy1}\\
&E_{y}^{(1)}\left(z\right)=E_{y}^{(1)}\left(0\right)\exp(z_1),\label{eq:Ex1}\\
&E_{z}^{(1)}\left(z\right)=-i\frac{k}{\beta^{(1)}}E_{y}^{(1)}\left(0\right)\exp(z_1),\label{eq:Ez1}
\end{align}

The solutions in the spacer region $j=2$ can be written up as:

\begin{align}
&H_{x}^{(2)}\left(z\right)=\frac{i\omega\varepsilon_0}{\beta^{(2)}}
\left[E_{y}^{(2)}(0){\cal F}_C(d_2-z_2)
-E_{y}^{(2)}(d) {\cal F}_C(z_2) \right],
 \\
&E_{y}^{(2)}(z)=
E_{y}^{(2)}(0) {\cal F}_S(d_2-z_2)+E_{y}^{(2)}(d)  {\cal F}_S(z_2) ,
\label{eq:Ex2}\\
& E_{z}^{(2)}(z) =\frac{ik}{\beta^{(2)}}\left[E_{y}^{(2)}(0){\cal F}_C(d_2-z_2)
-E_{y}^{(2)}(d){\cal F}_C(z_2) \right].
\label{eq:Ez2}
\end{align}

Finally, the solutions in the $j=3$ region, above graphene, read:

\begin {align}
&H_{x}^{(3)}(z) =i\frac{\omega\varepsilon_0}{\beta^{(2)}}E_{x}^{(3)}(d) e^{-(z_2-d_2)},\label{eq:Hy3}\\
&E_{y}^{(3)}\left(z\right)=E_{y}^{(3)}\left(d\right)e^{-(z_2-d_2)},\label{eq:Ex3}\\
&E_{z}^{(3)}\left(z\right)=i\frac{k}{\beta^{(2)}}E_{y}^{(3)}\left(d\right)e^{-(z_2-d_2)}.\label{eq:Ez3}
\end{align}
Here $\beta^{(1)}$ and $\beta^{(2)}$ are defined in Eqs.\,\eqref{eq:beta-1}. 

Using the same boundary condition equations as those in Eqs.\,\eqref{eq:bound-E-32-s}-\eqref{eq:bound-H-21-s} adapted for the TM polarization we obtain the linear homogeneous system of equations in the form
\begin{eqnarray}
\hspace{-0.5cm}
\left(\begin{array}{cc} 
 \frac{-1}{\beta^{(2)}\sinh(d_2)} & {\cal M}_{12} \\ 
 {\cal M}_{21} & \frac{1}{\beta^{(2)}\sinh(d_2)}
\end{array}
\right)
\left(\begin{array}{c} 
E_{y}^{(2)}(d) \\ E_{y}^{(2)}(0)
 \end{array}
\right)=
\left(\begin{array}{c} 
0 \\0  
\end{array}
\right),
\end{eqnarray}
where 

\begin{align}
&{\cal M}_{12}=\frac{1}{\beta^{(2)}\tanh(d_2)}+\frac{1}{\beta^{(1)}}\,,\\
&{\cal M}_{21}=
\frac{1}{\beta^{(2)}}+\frac{1}{\beta^{(2)}\tanh(d_2)}-\frac{\sigma\left(\omega\right)}{i\omega\varepsilon_0}\,.
\end{align}

Consequently, the dispersion relation can be represented as
\begin{eqnarray}
{\cal M}_{21}{\cal M}_{12}-\frac{1}{\left(\beta^{(2)}\right)^{2}\sinh^{2}(d_2)}=0\,.
\label{eq:dr-p}
\end{eqnarray}

\subsection{ Undressed SPP: the  $d\to\infty$  limit} 

\begin{figure}
	\includegraphics[width=8.5cm]{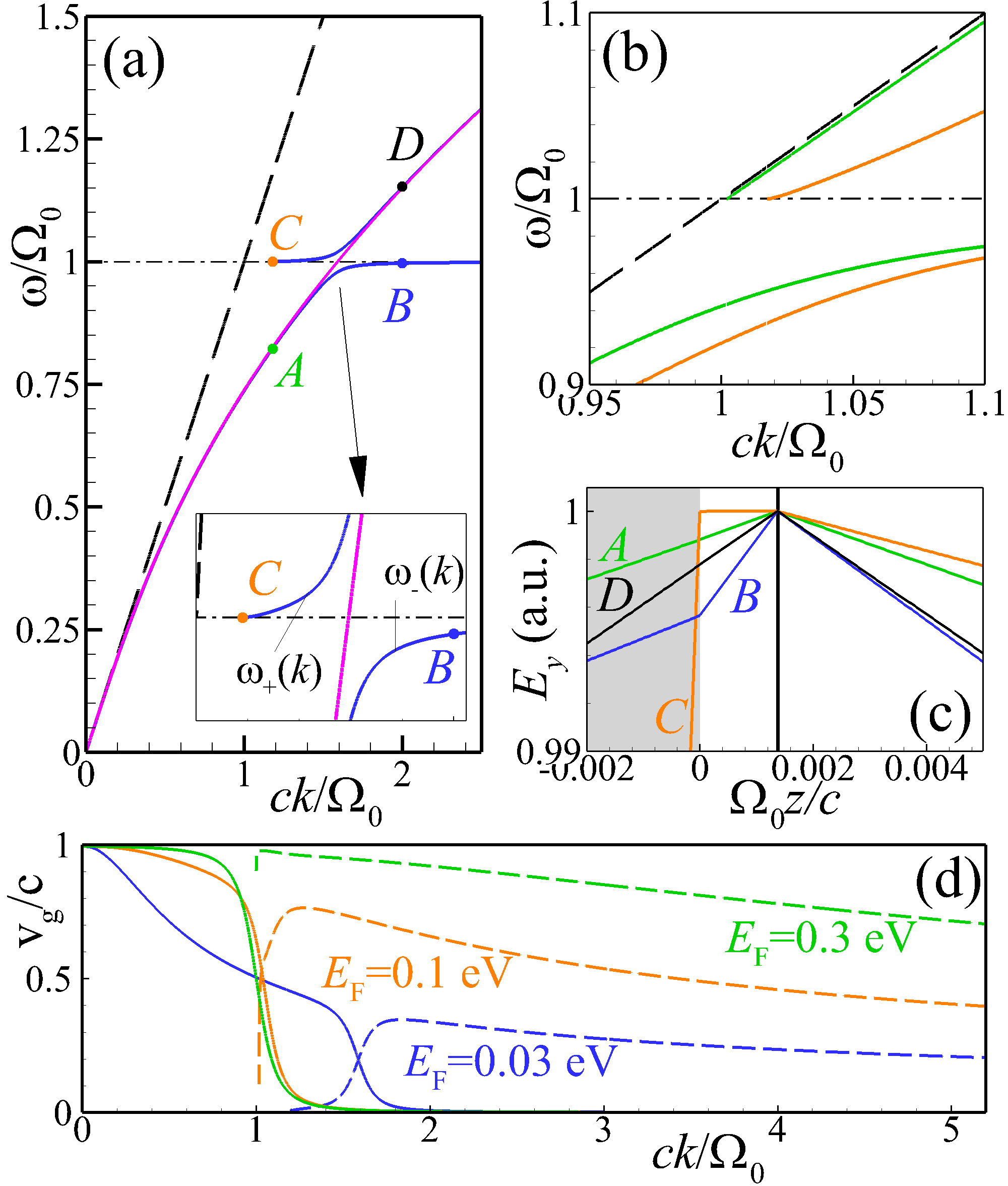}
	\caption{(a,b) Dispersion relations of TM-polarized SPMP in the graphene-antiferromagnet structure with $E_F=0.03\,$eV [solid blue lines in panel (a)], $E_F=0.1\,$eV [solid orange lines in panel (b)], and $E_F=0.3\,$eV [solid green lines in panel (b)]. For  comparison, the SPP dispersion relation in bare  graphene layer (with $E_F=0.03\,$eV)  is shown in panel (a) by solid pink line and the dashed black  line corresponds to the bare photon, $\omega=ck$, while the horizontal dash-and-dotted line corresponds to the AFMR frequency, $\omega=\Omega_{0}$. (c) Spatial profiles of the electric fields, corresponding to SPMP modes with $E_F=0.03\,$eV and $ck/\Omega_0=1.18$ (green line A and orange line C) and $ck/\Omega_0=2.0$ (blue line B and black line D). AF region is shadowed and graphene is at $z=0$. The field profiles are normalized to have the same magnitude on graphene. (d) Group velocity $v_g=d\omega/dk$ (in dimensionless units $v_g/c$) of the SPMP low-frequency [$\omega_-(k)$, solid lines] and high-frequency [$\omega_+(k)$, dashed lines] modes with $E_F=0.03\,$eV (blue lines), $E_F=0.1\,$eV (orange lines), and $E_F=0.4\,$eV (green lines). Other parameters of the structure are the same as those in Fig.\,\ref{fig:Spectrum-of-the-smp-graphene}.		\label{fig:TM-polar}}
\end{figure}

For infinite separation $d\to\infty$ between the graphene monolayer and the antiferromagnet the dispersion relation becomes:		
\begin{eqnarray}
\left[ \frac{2}{\beta^{(2)}}-\frac{\sigma\left(\omega\right)}{i\omega\varepsilon_0}\right]
\left[\frac{1}{\beta^{(2)}}+\frac{1}{\beta^{(1)}}\right]=0.\label{eq:dr-p-decoupled}
\end{eqnarray}
Setting equal to zero the first term in brackets in Eq.\,\eqref{eq:dr-p-decoupled} yields the dispersion relation of the SPPs in a free--standing graphene monolayer,
\begin{equation}
\beta^{(2)}=\sqrt{k^2-\frac{\omega^2}{c^2}}=\frac{2i\omega \epsilon_0 }{\sigma(\omega)}\,.
\label{eq:gr-spp}
\end{equation}
This dispersion curve is shown in Fig.\ref{fig:TM-polar}(a) for $E_F=0.03$ eV by solid pink line (this small Fermi energy is chosen for clarity of the figure). At low frequencies the dispersion relation \eqref{eq:gr-spp} can be expressed as
\begin{equation}
\omega=ck-\frac{\hbar^2}{8\alpha^2E_F^2}\left(ck\right)^3.\label{eq:gr-spp-approx}
\end{equation}
As a consequence, the dispersion curve \eqref{eq:gr-spp-approx} appears slightly below the light line.   

The second term in brackets in Eq.\,\eqref{eq:dr-p-decoupled} is always positive, so that it does not provide additional modes.
This situation, however changes when $d$ is finite.

\subsection{ SPP dressed by the antiferromagnet}

We now consider the effect of a  finite value of $d$  in  Eq.\,\eqref{eq:dr-p}.
We plot the spectrum of this hybrid modes -- SPMPs [solid blue lines in Fig.\,\ref{fig:TM-polar}].
We find that it consists of  two branches $\omega_-(k)$ and $\omega_+(k)$ with an anti-crossing between them [see inset in Fig. \ref{fig:TM-polar}],
which takes place in the vicinity of the AFMR frequency, $\omega=\Omega_0$. Away from the AFMR frequency  $\omega\lessgtr\Omega_0$, both modes follow the dispersion of the graphene 
SPP [like points A and D in Fig \ref{fig:TM-polar}(a)]. In the vicinity of the antiferromagnetic resonance the lower mode approaches asymptotically the AFMR frequency $\Omega_0$ as $k\to\infty$ [alike point B in Fig \ref{fig:TM-polar}(a), i.e., $\omega_-(\infty)=\Omega_0$], and the other one has the starting point of the spectrum at AFMR frequency, i.e. $\omega_+(k_+)=\Omega_0$ [point C in Fig \ref{fig:TM-polar}(a)].

The  strong enhancement of the density of states of the SPP at the bare (non-polaritonic)  AFMR frequency is a distinctive feature of the SPMPs, which are hybrid mode formed by the SPPs and the  bare magnons.
The magnitude of the SPMP mode  is quantified by:
\begin{enumerate}
\item  Absence of the energy gap between two branches, since $\omega=\Omega_0$ is both the maximum energy of the lower branch, and the minimum point of the upper branch; 
\item Presence of the infinite gap in momentum space at AFMR frequency $\Omega_0 =\omega_+(k_+)=\omega_-(\infty)$. This can be inferred by  group velocities of upper and lower branches, shown in Fig. \ref{fig:TM-polar}(d);
\item The asymmetry of the decay of the  $E_y$ component at two sides of the graphene sheet. 
\end{enumerate}

The electromagnetic field is, for all modes, predominantly concentrated nearby the graphene layer.
The distribution of the electric field depends strongly on the momentum $k$ and the SPMP branch. 
In Fig. \ref{fig:TM-polar}(c) we show $E_y(z)$, in units of $E_y(d)$, for 4 different SPMP modes, labelled with $A$,$B$,$C$ and $D$, shown in Fig. \ref{fig:TM-polar}(a). Modes $A$ and $D$ lie away from the anti-crossing of the branches and have a marked surface-plasmon character: their decay is the same at both sides ($z<d$ and $z>d$) of graphene, and their profile almost does not change at the AF surface ($z=0$). 
In contrast, modes $B$ and $C$ with frequencies nearby the AFMR frequency are asymmetric and change radically at the AF surface.

The evolution of the group velocity of the two branches as a function of $k$, for three values of $E_F$, shown in Fig.  \ref{fig:TM-polar}(d),  shows very clearly that the hybrid SPMP modes are combining the dispersive graphene SPP with a non-dispersive mode with $\omega=\Omega_0$. As the carrier density in graphene is varied, and thereby $E_F$,  SPMP dispersion is changed, and as a result, so is the value of $k$ at which the anticrossing takes place.

\section{Discussion and conclusions}
In this work we have investigated the electromagnetic properties of an antiferromagnetic insulator in the proximity of a graphene sheet. We have found two new types of hybrid polaritons that combine the electromagnetic field with the magnetization in the magnetic material and the  free carrier response of Dirac electrons in graphene:
\begin{enumerate}
\item A TE-polarized surface magnon-plasmon polariton (SMPP), propagating perpendicular to the direction of staggerred magnetization in the AF.  The group velocity of this mode becomes negative as $|E_F|$ is ramped up, resulting in a collective mode in the AF surface  whose propagation direction can be steered upon gating the graphene layer, located at a distance of a few hundred nanometers away.

\item A TM-polarized  surface plasmon-magnon polariton (SPMP), propagating along the staggerred magnetization direction,  which hybdridizes the surface-plasmon polariton in graphene and the bare magnons at the AF. 
\end{enumerate}

In both instances, 
a quantized theory of this new polaritons implies a
new type of  hybrid collective modes that combine of surface plasmons in graphene,  magnons in the antiferromagnets and the photon field. These new collective excitations have very exotic properties:
\begin{enumerate}
\item  They are a mixture of spin excitations (magnon), charged excitations (plasmon) and electromagnetic field (photon). The first term in the names that we attributed to these hybrid polaritons indicate the condensed matter excitation that primarily intercats with the field; it also determines its polarization (TE or TM). 
\item  They are extremely non-local, as they reside simultaneously at the graphene and the AF surface.
As a rule of thumb, the electromagnetic coupling between  these two layers survives as long as their separation $d$ is smaller than the wavelength of the  EM field at the relevant frequencies. Therefore, 
it survives to distances way above above 500 nanometers.
\item Their properties can be tuned by changing the carrier density in graphene.
\end{enumerate}

The recently discovered two-dimensional magnetic materials
\cite{GongZhang2017,HuangXu2017,mag2d-Miller2017-pt,mag2d-Hao2018-nat,FeiXu2018}  and the fabrication of Van der Waals heterostructures integrating 2D magnetic crystals with graphene and other non-magnetic 2D crystals \cite{WangMorpurgo2018,HuangXu2018,KleinJarillo2018,SongXu2018,JiangMak2018,JiangShan2018,SeylerXu2018}
opens up the  possibility of observing the same effects discussed in this work but in the context of van der Waals heterostructures. 
If the antiferromagnet is also 
metallic, then for frequencies below the plasma frequency, we would have a system exhibiting both negative permittivity and permeability functions.
This material would present intrinsic negative refraction.
In addition, the proximity of a graphene layer could be used for tuning the 
electromagnetic properties of the material. 

The prospects opened by 2D materials allow to envision many different arrangement of these systems leading to a new class of metamaterials with tunable electromagnetic properties promoted by the existence of magnetic order.


%

\section*{Acknowledgments}
Y. B., M. V. and N. M. R. P.  acknowledge support from the European Commission through the project "Graphene- Driven Revolutions in ICT and Beyond" (Ref. No. 785219), and the Portuguese Foundation for Science and Technology (FCT) in the framework of the Strategic Financing UID/FIS/04650/2013. Additionally, N. M. R. P. acknowledges  COMPETE2020, PORTUGAL2020, FEDER and the Portuguese Foundation for Science and Technology (FCT) through project PTDC/FIS-NAN/3668/2013 and FEDER and the portuguese Foundation for Science and Technology (FCT) through project
POCI-01-0145-FEDER-028114.
G. A. Farias  acknowledge support from the Conselho Nacional de Desenvolvimento Cient\'{i}fico e Tecnol\'ogico (CNPq) of Brazil.
J. F.-R. acknowledge financial support
from FCT for the P2020-PTDC/FIS-NAN/4662/2014, the P2020-PTDC/FIS-NAN/3668/2014 and the UTAPEXPL/NTec/0046/2017 projects, as well as Generalitat
Valenciana funding Prometeo2017/139 and MINECO Spain (Grant No. MAT2016-78625-C2).

\appendix

\section{Transverse bulk waves \label{app}}

Let us imagine that antiferromagnetic medium characterized by the magnetic permeability tensor (\ref{eq:tensor-mu}) occupies all the space $-\infty<z<\infty$. In this case wave propagation is governed by Maxwell equations (\ref{eq:Maxwell-j-1}), which solutions we will seek in the form of travelling waves $\boldsymbol{\mathcal{E}}(\mathbf{r},t)=\mathbf{E}\exp(i\mathbf{kr}-i\omega t)$,
$\boldsymbol{\mathcal{H}}(\mathbf{r},t)=\mathbf{H}\exp(i\mathbf{kr}-i\omega t)$, propagating in arbitrary direction $k$ with amplitudes of electromagnetic field $\mathbf{E}$, $\mathbf{H}$. Under this assumption, jointly with the constitutive relations $\boldsymbol{\mathcal{D}}(\mathbf{r},t)=\varepsilon_0\mathbf{E}\exp(i\mathbf{kr}-i\omega t)$,
$\boldsymbol{\mathcal{B}}(\mathbf{r},t)=\mu_0\hat{\boldsymbol{\mu}}\left(\omega\right)\mathbf{H}\exp(i\mathbf{kr}-i\omega t)$
Maxwell equations (\ref{eq:Maxwell-j-1}) will be rewritten as 
\begin{eqnarray}
\mathbf{k}\times\mathbf{H}=-\omega\varepsilon_0\mathbf{E},\label{eq:Maxwell-amp-1}\\
\mathbf{k}\times\mathbf{E}=\omega\mu_0\hat{\boldsymbol{\mu}}\left(\omega\right)\mathbf{H},\label{eq:Maxwell-amp-2}\\
i\mathbf{k}\cdot\mathbf{E}=0,\label{eq:Maxwell-amp-3}\\
i\mathbf{k}\cdot\hat{\boldsymbol{\mu}}\left(\omega\right)\mathbf{H}=0.\label{eq:Maxwell-amp-4}
\end{eqnarray}
Notice that in this Appendix we omit index $j$ for brevity.
If we apply operator $\mathbf{k}\times$ to Eq.\,\eqref{eq:Maxwell-amp-1}
and use Eq.\,\eqref{eq:Maxwell-amp-2}, we have
\begin{eqnarray}
\mathbf{k}\times(\mathbf{k}\times\mathbf{H})=\mathbf{k}\left(\mathbf{k}\cdot\mathbf{H}\right)-k^{2}\mathbf{H}\nonumber\\
=-\left(\frac{\omega}{c}\right)^{2}\hat{\boldsymbol{\mu}}\left(\omega\right)\mathbf{H}.\label{eq:Helmholtz-antiferromagnet}
\end{eqnarray}
For the transverse waves the wavevector $\mathbf{k}$ should be orthogonal
to the magnetic field $\mathbf{H}$, i.e. $\left(\mathbf{k}\cdot\mathbf{H}\right)=0$.
Further this wave will be referred to as TM-polarized bulk polariton.
Simultaneously with Eq.\,\eqref{eq:Maxwell-amp-4} this condition
can be satisfied only if $H_{y}\equiv0$. In this case the Helmholtz equation (\ref{eq:Helmholtz-antiferromagnet}) for components of the magnetic field $H_x$ and $H_z$ will be rewritten as

\begin{align}
	&\left[k^{2}-\left(\frac{\omega}{c}\right)^{2}\mu\left(\omega\right)\right]H_{x}=0=\beta_1 H_x,\\
	&\left[k^{2}-\left(\frac{\omega}{c}\right)^{2}\mu\left(\omega\right)\right]H_{z}=0=\beta_1 H_z.
\end{align}

For nonzero amplitudes this system of equation will have solution only when condition $\beta_1=0$ is met, thus Eq.\,(\ref{eq:bulk-dr}) determines the dispersion relation of bulk waves. If the wavevector is represented in spherical coordinates as 
\begin{eqnarray}
\mathbf{k}=\frac{\omega}{c}\sqrt{\mu\left(\omega\right)}\left(\mathbf{u}_x\cos\varphi\sin\theta+\mathbf{u}_y\cos\theta\right.\nonumber\\
\left.+\mathbf{u}_z\sin\varphi\sin\theta\right),
\end{eqnarray}
the respective components of the magnetic field will be $\mathbf{H}=H\left(-\mathbf{u}_x\sin\varphi+\mathbf{u}_z\cos\varphi\right)$. In these equations $\theta$ is the polar angle between the y-axis and wavevector, and $\varphi$ is the azimuthal angle in plane $xz$.
The electric field is also perpendicular
to the direction of the propagation owing the Eq.\eqref{eq:Maxwell-amp-3}.
The components of the respective electric field can be defined from
Eq.\,\eqref{eq:Maxwell-amp-1}
\begin{eqnarray}
	\mathbf{E}=H\sqrt{\frac{\mu_0\mu_a\left(\omega\right)}{\varepsilon_0}}\\
	\times \left[-\mathbf{u}_{x}\cos\varphi\cos\theta+\mathbf{u}_{y}\sin\theta-\mathbf{u}_{z}\sin\varphi\cos\theta\right].\nonumber
\end{eqnarray}
Notice that in this representation the electric field is perpendicular to magnetic field, $\mathbf{E}\perp\mathbf{H}$, what follows from the scalar product $\mathbf{E}\cdot\mathbf{H}=0$. 

Eq.\,\eqref{eq:bulk-dr} have two solution for $\omega$ -- acoustic $\omega_a$ and optical $\omega_o$ modes 
\begin{align*}
\omega_{a} & =\sqrt{f(k)-\sqrt{f(k)^2-c^{2}k^{2}\Omega_{0}^{2}}},\\
\omega_{o} & =\sqrt{f(k)+\sqrt{f(k)^2-c^{2}k^{2}\Omega_{0}^{2}}},
\end{align*}
with
\begin{equation}
f(k)=\frac{c^{2}k^{2}+\Omega_{0}^{2}}{2}+\Omega_{s}^{2}.
\end{equation}
This is dissimilar
to the case of a metal described by a Drude optical response, where only one transverse
bulk mode exists.
Spectra of these two bulk TM-polarized magnon-polariton modes are depicted in Fig.\,\ref{fig:Schematic-dispersion-relation} by dashed red lines.
The spectrum of the acoustic mode starts at zero frequency and in at long-wavelength limit $k\rightarrow0$ is described by the approximate expression as
\begin{equation}
\omega_{a} \approx kc\sqrt{\frac{\Omega_{0}^{2}}{\Omega_{0}^{2}+2\Omega_{s}^{2}}}.
\end{equation}
In short-wavelength limit $k\rightarrow\infty$ the dispersion curve of acoustic mode asymptotically approaches the frequency $\Omega_0$ as  
\begin{equation}
\omega_{a} \approx\Omega_{0}-\frac{\Omega_{0}\Omega_{s}^{2}}{(ck)^{2}}.
\end{equation}
It should be underlined that the spectrum of the acoustic mode is located at the right of the light line $\omega=ck$ (depicted by dashed black line in Fig.\,\ref{fig:Spectrum-of-the-smp-graphene}). This fact means that phase velocity of the acoustic mode, $\omega_a/k$ is smaller than the velocity of light in vacuum $c$ for all values of the wavevector $k$.

The optical mode spectrum starts at the frequency $\sqrt{\Omega_{0}^{2}+2\Omega_{s}^{2}}$, and in the limit $k\to 0$ its approximate dispersion relation can be represented as
\begin{equation}
\omega_{o} \approx\sqrt{\Omega_{0}^{2}+2\Omega_{s}^{2}}+\frac{c^{2}k^{2}\Omega_{s}^{2}}{(\Omega_{0}^{2}+2\Omega_{s}^{2})^{3/2}},
\end{equation}
while in the limit $k\to\infty$ the optical mode's approximate dispersion relation is 
\begin{equation}
\omega_{o} \approx kc+\frac{\Omega_{s}^{2}}{ck}.
\end{equation}
Thus, at large values of wavevector $k$, the optical mode spectrum asymptotically approaches light line $\omega=ck$. Contrary to the acoustic mode, the optical one is characterized by phase velocity larger than the velocity of light in vacuum $c$, and its spectrum is located at the left of the light line. 

It is interesting that no  bulk magnon polariton mode exists in the frequency range $\Omega_0<\omega<\sqrt{\Omega_{0}^{2}+2\Omega_{s}^{2}}$ -- between the highest frequency of the acoustic mode and lowest frequency of the optical mode. Notice, that this gap is characterized by the negative values of $\mu_a(\omega)<0$.

\bibliographystyle{apsrev4-1}

\begin{thebibliography}{34}%
\makeatletter
\providecommand \@ifxundefined [1]{%
 \@ifx{#1\undefined}
}%
\providecommand \@ifnum [1]{%
 \ifnum #1\expandafter \@firstoftwo
 \else \expandafter \@secondoftwo
 \fi
}%
\providecommand \@ifx [1]{%
 \ifx #1\expandafter \@firstoftwo
 \else \expandafter \@secondoftwo
 \fi
}%
\providecommand \natexlab [1]{#1}%
\providecommand \enquote  [1]{``#1''}%
\providecommand \bibnamefont  [1]{#1}%
\providecommand \bibfnamefont [1]{#1}%
\providecommand \citenamefont [1]{#1}%
\providecommand \href@noop [0]{\@secondoftwo}%
\providecommand \href [0]{\begingroup \@sanitize@url \@href}%
\providecommand \@href[1]{\@@startlink{#1}\@@href}%
\providecommand \@@href[1]{\endgroup#1\@@endlink}%
\providecommand \@sanitize@url [0]{\catcode `\\12\catcode `\$12\catcode
  `\&12\catcode `\#12\catcode `\^12\catcode `\_12\catcode `\%12\relax}%
\providecommand \@@startlink[1]{}%
\providecommand \@@endlink[0]{}%
\providecommand \url  [0]{\begingroup\@sanitize@url \@url }%
\providecommand \@url [1]{\endgroup\@href {#1}{\urlprefix }}%
\providecommand \urlprefix  [0]{URL }%
\providecommand \Eprint [0]{\href }%
\providecommand \doibase [0]{http://dx.doi.org/}%
\providecommand \selectlanguage [0]{\@gobble}%
\providecommand \bibinfo  [0]{\@secondoftwo}%
\providecommand \bibfield  [0]{\@secondoftwo}%
\providecommand \translation [1]{[#1]}%
\providecommand \BibitemOpen [0]{}%
\providecommand \bibitemStop [0]{}%
\providecommand \bibitemNoStop [0]{.\EOS\space}%
\providecommand \EOS [0]{\spacefactor3000\relax}%
\providecommand \BibitemShut  [1]{\csname bibitem#1\endcsname}%
\let\auto@bib@innerbib\@empty
\bibitem [{\citenamefont {Keffer}\ and\ \citenamefont
  {Kittel}(1952)}]{antiferro-Keffer1952-pr}%
  \BibitemOpen
  \bibfield  {author} {\bibinfo {author} {\bibfnamefont {F.}~\bibnamefont
  {Keffer}}\ and\ \bibinfo {author} {\bibfnamefont {C.}~\bibnamefont
  {Kittel}},\ }\href {\doibase 10.1103/PhysRev.85.329} {\bibfield  {journal}
  {\bibinfo  {journal} {Physical Review}\ }\textbf {\bibinfo {volume} {85}},\
  \bibinfo {pages} {329} (\bibinfo {year} {1952})}\BibitemShut {NoStop}%
\bibitem [{\citenamefont {Camley}\ and\ \citenamefont
  {Mills}(1982)}]{smp-Camley1982-prb}%
  \BibitemOpen
  \bibfield  {author} {\bibinfo {author} {\bibfnamefont {R.~E.}\ \bibnamefont
  {Camley}}\ and\ \bibinfo {author} {\bibfnamefont {D.~L.}\ \bibnamefont
  {Mills}},\ }\href {\doibase 10.1103/PhysRevB.26.1280} {\bibfield  {journal}
  {\bibinfo  {journal} {Phys. Rev. B}\ }\textbf {\bibinfo {volume} {26}},\
  \bibinfo {pages} {1280} (\bibinfo {year} {1982})}\BibitemShut {NoStop}%
\bibitem [{\citenamefont {Jensen}\ \emph {et~al.}(1995)\citenamefont {Jensen},
  \citenamefont {Parker}, \citenamefont {Abraha},\ and\ \citenamefont
  {Tilley}}]{smp-exp-Jensen1995-prl}%
  \BibitemOpen
  \bibfield  {author} {\bibinfo {author} {\bibfnamefont {M.~R.~F.}\
  \bibnamefont {Jensen}}, \bibinfo {author} {\bibfnamefont {T.~J.}\
  \bibnamefont {Parker}}, \bibinfo {author} {\bibfnamefont {K.}~\bibnamefont
  {Abraha}}, \ and\ \bibinfo {author} {\bibfnamefont {D.~R.}\ \bibnamefont
  {Tilley}},\ }\href {\doibase 10.1103/PhysRevLett.75.3756} {\bibfield
  {journal} {\bibinfo  {journal} {Phys. Rev. Lett.}\ }\textbf {\bibinfo
  {volume} {75}},\ \bibinfo {pages} {3756} (\bibinfo {year}
  {1995})}\BibitemShut {NoStop}%
\bibitem [{\citenamefont {Padilla}\ \emph {et~al.}(2006)\citenamefont
  {Padilla}, \citenamefont {Basov},\ and\ \citenamefont
  {Smith}}]{metamaterial-Padilla2006-mt}%
  \BibitemOpen
  \bibfield  {author} {\bibinfo {author} {\bibfnamefont {W.~J.}\ \bibnamefont
  {Padilla}}, \bibinfo {author} {\bibfnamefont {D.~N.}\ \bibnamefont {Basov}},
  \ and\ \bibinfo {author} {\bibfnamefont {D.~R.}\ \bibnamefont {Smith}},\
  }\href {\doibase 10.1016/S1369-7021(06)71573-5} {\bibfield  {journal}
  {\bibinfo  {journal} {Materials Today}\ }\textbf {\bibinfo {volume} {9}},\
  \bibinfo {pages} {28} (\bibinfo {year} {2006})}\BibitemShut {NoStop}%
\bibitem [{\citenamefont {Kaganov}\ and\ \citenamefont
  {Tsukernik}(1958)}]{antiferro-Kaganov1958-jetp}%
  \BibitemOpen
  \bibfield  {author} {\bibinfo {author} {\bibfnamefont {M.}~\bibnamefont
  {Kaganov}}\ and\ \bibinfo {author} {\bibfnamefont {V.}~\bibnamefont
  {Tsukernik}},\ }\href
  {http://www.jetp.ac.ru/cgi-bin/e/index/e/7/2/p361?a=list} {\bibfield
  {journal} {\bibinfo  {journal} {Journal of Experimental and Theoretical
  Physics}\ }\textbf {\bibinfo {volume} {7}},\ \bibinfo {pages} {361} (\bibinfo
  {year} {1958})}\BibitemShut {NoStop}%
\bibitem [{\citenamefont {Kaganov}\ and\ \citenamefont
  {Tsukernik}(1962)}]{antiferro-Kaganov1962-jetp}%
  \BibitemOpen
  \bibfield  {author} {\bibinfo {author} {\bibfnamefont {M.}~\bibnamefont
  {Kaganov}}\ and\ \bibinfo {author} {\bibfnamefont {V.}~\bibnamefont
  {Tsukernik}},\ }\href
  {http://www.jetp.ac.ru/cgi-bin/e/index/e/14/1/p192?a=list} {\bibfield
  {journal} {\bibinfo  {journal} {Journal of Experimental and Theoretical
  Physics}\ }\textbf {\bibinfo {volume} {14}},\ \bibinfo {pages} {192}
  (\bibinfo {year} {1962})}\BibitemShut {NoStop}%
\bibitem [{\citenamefont {Mills}\ and\ \citenamefont
  {Burstein}(1974)}]{smp-Mills1974-rpp}%
  \BibitemOpen
  \bibfield  {author} {\bibinfo {author} {\bibfnamefont {D.~L.}\ \bibnamefont
  {Mills}}\ and\ \bibinfo {author} {\bibfnamefont {E.}~\bibnamefont
  {Burstein}},\ }\href {\doibase 10.1088/0034-4885/37/7/001} {\bibfield
  {journal} {\bibinfo  {journal} {Rep. Prog. Phys.}\ }\textbf {\bibinfo
  {volume} {37}},\ \bibinfo {pages} {817} (\bibinfo {year} {1974})}\BibitemShut
  {NoStop}%
\bibitem [{\citenamefont {Almeida}\ \emph {et~al.}(1993)\citenamefont
  {Almeida}, \citenamefont {Farias}, \citenamefont {Oliveira},\ and\
  \citenamefont {Vasconcelos}}]{antiferro-Almeida1993-prb}%
  \BibitemOpen
  \bibfield  {author} {\bibinfo {author} {\bibfnamefont {N.~S.}\ \bibnamefont
  {Almeida}}, \bibinfo {author} {\bibfnamefont {G.~A.}\ \bibnamefont {Farias}},
  \bibinfo {author} {\bibfnamefont {N.~T.}\ \bibnamefont {Oliveira}}, \ and\
  \bibinfo {author} {\bibfnamefont {E.~F.}\ \bibnamefont {Vasconcelos}},\
  }\href {\doibase 10.1103/PhysRevB.48.9839} {\bibfield  {journal} {\bibinfo
  {journal} {Physical Review B}\ }\textbf {\bibinfo {volume} {48}},\ \bibinfo
  {pages} {9839} (\bibinfo {year} {1993})}\BibitemShut {NoStop}%
\bibitem [{\citenamefont {Tarkhanyan}\ and\ \citenamefont
  {Niarchos}(2009)}]{smp-ni-Tarkhanyan2009-pssb}%
  \BibitemOpen
  \bibfield  {author} {\bibinfo {author} {\bibfnamefont {R.~H.}\ \bibnamefont
  {Tarkhanyan}}\ and\ \bibinfo {author} {\bibfnamefont {D.~G.}\ \bibnamefont
  {Niarchos}},\ }\href {\doibase 10.1002/pssb.200945071} {\bibfield  {journal}
  {\bibinfo  {journal} {physica status solidi (b)}\ }\textbf {\bibinfo {volume}
  {246}},\ \bibinfo {pages} {1939} (\bibinfo {year} {2009})}\BibitemShut
  {NoStop}%
\bibitem [{\citenamefont {Koppens}\ \emph {et~al.}(2011)\citenamefont
  {Koppens}, \citenamefont {Chang},\ and\ \citenamefont {{Garc\'{i}a de
  Abajo}}}]{spp-gr-rev-Koppens2011-nl}%
  \BibitemOpen
  \bibfield  {author} {\bibinfo {author} {\bibfnamefont {F.~H.~L.}\
  \bibnamefont {Koppens}}, \bibinfo {author} {\bibfnamefont {D.~E.}\
  \bibnamefont {Chang}}, \ and\ \bibinfo {author} {\bibfnamefont {F.~J.}\
  \bibnamefont {{Garc\'{i}a de Abajo}}},\ }\href {\doibase 10.1021/nl201771h}
  {\bibfield  {journal} {\bibinfo  {journal} {Nano Letters}\ }\textbf {\bibinfo
  {volume} {11}},\ \bibinfo {pages} {3370} (\bibinfo {year} {2011})},\ \Eprint
  {http://arxiv.org/abs/1104.2068} {arXiv:1104.2068} \BibitemShut {NoStop}%
\bibitem [{\citenamefont {Nikitin}\ \emph {et~al.}(2011)\citenamefont
  {Nikitin}, \citenamefont {Guinea}, \citenamefont {Garc{\'{i}}a-Vidal},\ and\
  \citenamefont {Mart{\'{i}}n-Moreno}}]{spp-gr-Nikitin2011-prb}%
  \BibitemOpen
  \bibfield  {author} {\bibinfo {author} {\bibfnamefont {A.~Y.}\ \bibnamefont
  {Nikitin}}, \bibinfo {author} {\bibfnamefont {F.}~\bibnamefont {Guinea}},
  \bibinfo {author} {\bibfnamefont {F.~J.}\ \bibnamefont {Garc{\'{i}}a-Vidal}},
  \ and\ \bibinfo {author} {\bibfnamefont {L.}~\bibnamefont
  {Mart{\'{i}}n-Moreno}},\ }\href {\doibase 10.1103/PhysRevB.84.195446}
  {\bibfield  {journal} {\bibinfo  {journal} {Physical Review B}\ }\textbf
  {\bibinfo {volume} {84}},\ \bibinfo {pages} {195446} (\bibinfo {year}
  {2011})}\BibitemShut {NoStop}%
\bibitem [{\citenamefont {Brar}\ \emph {et~al.}(2014)\citenamefont {Brar},
  \citenamefont {Jang}, \citenamefont {Sherrott}, \citenamefont {Kim},
  \citenamefont {Lopez}, \citenamefont {Kim}, \citenamefont {Choi},\ and\
  \citenamefont {Atwater}}]{spp-gr-phon-Brar2014-nl}%
  \BibitemOpen
  \bibfield  {author} {\bibinfo {author} {\bibfnamefont {V.~W.}\ \bibnamefont
  {Brar}}, \bibinfo {author} {\bibfnamefont {M.~S.}\ \bibnamefont {Jang}},
  \bibinfo {author} {\bibfnamefont {M.}~\bibnamefont {Sherrott}}, \bibinfo
  {author} {\bibfnamefont {S.}~\bibnamefont {Kim}}, \bibinfo {author}
  {\bibfnamefont {J.~J.}\ \bibnamefont {Lopez}}, \bibinfo {author}
  {\bibfnamefont {L.~B.}\ \bibnamefont {Kim}}, \bibinfo {author} {\bibfnamefont
  {M.}~\bibnamefont {Choi}}, \ and\ \bibinfo {author} {\bibfnamefont
  {H.}~\bibnamefont {Atwater}},\ }\href {\doibase 10.1021/nl501096s} {\bibfield
   {journal} {\bibinfo  {journal} {Nano Letters}\ }\textbf {\bibinfo {volume}
  {14}},\ \bibinfo {pages} {3876} (\bibinfo {year} {2014})}\BibitemShut
  {NoStop}%
\bibitem [{\citenamefont {Liu}\ and\ \citenamefont
  {Willis}(2010)}]{spp-gr-phon-Liu2010-prb}%
  \BibitemOpen
  \bibfield  {author} {\bibinfo {author} {\bibfnamefont {Y.}~\bibnamefont
  {Liu}}\ and\ \bibinfo {author} {\bibfnamefont {R.~F.}\ \bibnamefont
  {Willis}},\ }\href {\doibase 10.1103/PhysRevB.81.081406} {\bibfield
  {journal} {\bibinfo  {journal} {Physical Review B}\ }\textbf {\bibinfo
  {volume} {81}},\ \bibinfo {pages} {081406} (\bibinfo {year}
  {2010})}\BibitemShut {NoStop}%
\bibitem [{\citenamefont {Luxmoore}\ \emph {et~al.}(2014)\citenamefont
  {Luxmoore}, \citenamefont {Gan}, \citenamefont {Liu}, \citenamefont
  {Valmorra}, \citenamefont {Li}, \citenamefont {Faist},\ and\ \citenamefont
  {Nash}}]{spp-gr-phon-Luxmoore2014-acsphot}%
  \BibitemOpen
  \bibfield  {author} {\bibinfo {author} {\bibfnamefont {I.~J.}\ \bibnamefont
  {Luxmoore}}, \bibinfo {author} {\bibfnamefont {C.~H.}\ \bibnamefont {Gan}},
  \bibinfo {author} {\bibfnamefont {P.~Q.}\ \bibnamefont {Liu}}, \bibinfo
  {author} {\bibfnamefont {F.}~\bibnamefont {Valmorra}}, \bibinfo {author}
  {\bibfnamefont {P.}~\bibnamefont {Li}}, \bibinfo {author} {\bibfnamefont
  {J.}~\bibnamefont {Faist}}, \ and\ \bibinfo {author} {\bibfnamefont {G.~R.}\
  \bibnamefont {Nash}},\ }\href {\doibase 10.1021/ph500233s} {\bibfield
  {journal} {\bibinfo  {journal} {ACS Photonics}\ }\textbf {\bibinfo {volume}
  {1}},\ \bibinfo {pages} {1151} (\bibinfo {year} {2014})}\BibitemShut
  {NoStop}%
\bibitem [{\citenamefont {Fei}\ \emph {et~al.}(2011)\citenamefont {Fei},
  \citenamefont {Andreev}, \citenamefont {Bao}, \citenamefont {Zhang},
  \citenamefont {{S McLeod}}, \citenamefont {Wang}, \citenamefont {Stewart},
  \citenamefont {Zhao}, \citenamefont {Dominguez}, \citenamefont {Thiemens},
  \citenamefont {Fogler}, \citenamefont {Tauber}, \citenamefont {Castro-Neto},
  \citenamefont {Lau}, \citenamefont {Keilmann},\ and\ \citenamefont
  {Basov}}]{spp-gr-phon-Fei2011-nl}%
  \BibitemOpen
  \bibfield  {author} {\bibinfo {author} {\bibfnamefont {Z.}~\bibnamefont
  {Fei}}, \bibinfo {author} {\bibfnamefont {G.~O.}\ \bibnamefont {Andreev}},
  \bibinfo {author} {\bibfnamefont {W.}~\bibnamefont {Bao}}, \bibinfo {author}
  {\bibfnamefont {L.~M.}\ \bibnamefont {Zhang}}, \bibinfo {author}
  {\bibfnamefont {A.}~\bibnamefont {{S McLeod}}}, \bibinfo {author}
  {\bibfnamefont {C.}~\bibnamefont {Wang}}, \bibinfo {author} {\bibfnamefont
  {M.~K.}\ \bibnamefont {Stewart}}, \bibinfo {author} {\bibfnamefont
  {Z.}~\bibnamefont {Zhao}}, \bibinfo {author} {\bibfnamefont {G.}~\bibnamefont
  {Dominguez}}, \bibinfo {author} {\bibfnamefont {M.}~\bibnamefont {Thiemens}},
  \bibinfo {author} {\bibfnamefont {M.~M.}\ \bibnamefont {Fogler}}, \bibinfo
  {author} {\bibfnamefont {M.~J.}\ \bibnamefont {Tauber}}, \bibinfo {author}
  {\bibfnamefont {A.~H.}\ \bibnamefont {Castro-Neto}}, \bibinfo {author}
  {\bibfnamefont {C.~N.}\ \bibnamefont {Lau}}, \bibinfo {author} {\bibfnamefont
  {F.}~\bibnamefont {Keilmann}}, \ and\ \bibinfo {author} {\bibfnamefont
  {D.~N.}\ \bibnamefont {Basov}},\ }\href {\doibase 10.1021/nl202362d}
  {\bibfield  {journal} {\bibinfo  {journal} {Nano letters}\ ,\ \bibinfo
  {pages} {4701}} (\bibinfo {year} {2011})}\BibitemShut {NoStop}%
\bibitem [{\citenamefont {Kumar}\ \emph {et~al.}(2015)\citenamefont {Kumar},
  \citenamefont {Low}, \citenamefont {Fung}, \citenamefont {Avouris},\ and\
  \citenamefont {Fang}}]{kumar2015}%
  \BibitemOpen
  \bibfield  {author} {\bibinfo {author} {\bibfnamefont {A.}~\bibnamefont
  {Kumar}}, \bibinfo {author} {\bibfnamefont {T.}~\bibnamefont {Low}}, \bibinfo
  {author} {\bibfnamefont {K.~H.}\ \bibnamefont {Fung}}, \bibinfo {author}
  {\bibfnamefont {P.}~\bibnamefont {Avouris}}, \ and\ \bibinfo {author}
  {\bibfnamefont {N.~X.}\ \bibnamefont {Fang}},\ }\href@noop {} {\bibfield
  {journal} {\bibinfo  {journal} {Nano letters}\ }\textbf {\bibinfo {volume}
  {15}},\ \bibinfo {pages} {3172} (\bibinfo {year} {2015})}\BibitemShut
  {NoStop}%
\bibitem [{\citenamefont {Guti\'errez-Rubio}\ \emph {et~al.}(2013)\citenamefont
  {Guti\'errez-Rubio}, \citenamefont {Stauber},\ and\ \citenamefont
  {Guinea}}]{Stauber2013}%
  \BibitemOpen
  \bibfield  {author} {\bibinfo {author} {\bibfnamefont {A.}~\bibnamefont
  {Guti\'errez-Rubio}}, \bibinfo {author} {\bibfnamefont {T.}~\bibnamefont
  {Stauber}}, \ and\ \bibinfo {author} {\bibfnamefont {F.}~\bibnamefont
  {Guinea}},\ }\href@noop {} {\bibfield  {journal} {\bibinfo  {journal} {J.
  Optics}\ }\textbf {\bibinfo {volume} {15}},\ \bibinfo {pages} {114005}
  (\bibinfo {year} {2013})}\BibitemShut {NoStop}%
\bibitem [{Not()}]{Note2}%
  \BibitemOpen
  \href@noop {} {}\bibinfo {note} {The inclusion of losses is straightforward
  but complicates the analysis and is, therefore, ignored.}\BibitemShut {Stop}%
\bibitem [{Note1()}]{Note1}%
  \BibitemOpen
  \bibinfo {note} {If the interband transitions are taken into account in the
  expression of graphene's conductivity, its imaginary part can be negative. In
  this case graphene is able to sustain s-polarized surface
  plasmon-polaritons\cite {gr-spp-te-Mikhailov2007-prl}, which dispersion
  relation is described by the term in first braces of Eq.\protect \tmspace
  +\thinmuskip {.1667em}\protect \textup {\hbox {\mathsurround \z@ \protect
  \normalfont (\ignorespaces \ref {eq:dr-s-decoupled}\unskip \@@italiccorr )}}.
  Nevertheless, their typical frequencies lies in the vicinity of double
  graphene's Fermi energy $\omega \sim 2E_F$, i.e. significantly higher than
  the antiferromagnet resonance frequency $\Omega _0$.}\BibitemShut {Stop}%
\bibitem [{\citenamefont {Mac\^{e}do}(2017)}]{antiferro-Macedo2017}%
  \BibitemOpen
  \bibfield  {author} {\bibinfo {author} {\bibfnamefont {R.}~\bibnamefont
  {Mac\^{e}do}},\ }\enquote {\bibinfo {title} {Chapter two - tunable hyperbolic
  media: Magnon-polaritons in canted antiferromagnets},}\ \ (\bibinfo
  {publisher} {Academic Press},\ \bibinfo {year} {2017})\ pp.\ \bibinfo {pages}
  {91 -- 155}\BibitemShut {NoStop}%
\bibitem [{\citenamefont {Jablan}\ \emph {et~al.}(2009)\citenamefont {Jablan},
  \citenamefont {Buljan},\ and\ \citenamefont
  {Solja{\v{c}}i{\'{c}}}}]{gr-spp-Jablan2009-prb}%
  \BibitemOpen
  \bibfield  {author} {\bibinfo {author} {\bibfnamefont {M.}~\bibnamefont
  {Jablan}}, \bibinfo {author} {\bibfnamefont {H.}~\bibnamefont {Buljan}}, \
  and\ \bibinfo {author} {\bibfnamefont {M.}~\bibnamefont
  {Solja{\v{c}}i{\'{c}}}},\ }\href {\doibase 10.1103/PhysRevB.80.245435}
  {\bibfield  {journal} {\bibinfo  {journal} {Physical Review B}\ }\textbf
  {\bibinfo {volume} {80}} (\bibinfo {year} {2009}),\
  10.1103/PhysRevB.80.245435}\BibitemShut {NoStop}%
\bibitem [{\citenamefont {Bludov}\ \emph {et~al.}(2013)\citenamefont {Bludov},
  \citenamefont {Ferreira}, \citenamefont {Peres},\ and\ \citenamefont
  {Vasilevskiy}}]{spp-gr-rev-Bludov2013-ijmfb}%
  \BibitemOpen
  \bibfield  {author} {\bibinfo {author} {\bibfnamefont {Y.~V.}\ \bibnamefont
  {Bludov}}, \bibinfo {author} {\bibfnamefont {A.}~\bibnamefont {Ferreira}},
  \bibinfo {author} {\bibfnamefont {N.~M.~R.}\ \bibnamefont {Peres}}, \ and\
  \bibinfo {author} {\bibfnamefont {M.~I.}\ \bibnamefont {Vasilevskiy}},\
  }\href {\doibase 10.1142/S0217979213410014} {\bibfield  {journal} {\bibinfo
  {journal} {International Journal of Modern Physics B}\ }\textbf {\bibinfo
  {volume} {27}},\ \bibinfo {pages} {1341001} (\bibinfo {year}
  {2013})}\BibitemShut {NoStop}%
\bibitem [{\citenamefont {Gong}\ \emph {et~al.}(2017)\citenamefont {Gong},
  \citenamefont {Li}, \citenamefont {Li}, \citenamefont {Ji}, \citenamefont
  {Stern}, \citenamefont {Xia}, \citenamefont {Cao}, \citenamefont {Bao},
  \citenamefont {Wang}, \citenamefont {Wang}, \citenamefont {Qiu},
  \citenamefont {Cava}, \citenamefont {Louie}, \citenamefont {Xia},\ and\
  \citenamefont {Zhang}}]{GongZhang2017}%
  \BibitemOpen
  \bibfield  {author} {\bibinfo {author} {\bibfnamefont {C.}~\bibnamefont
  {Gong}}, \bibinfo {author} {\bibfnamefont {L.}~\bibnamefont {Li}}, \bibinfo
  {author} {\bibfnamefont {Z.}~\bibnamefont {Li}}, \bibinfo {author}
  {\bibfnamefont {H.}~\bibnamefont {Ji}}, \bibinfo {author} {\bibfnamefont
  {A.}~\bibnamefont {Stern}}, \bibinfo {author} {\bibfnamefont
  {Y.}~\bibnamefont {Xia}}, \bibinfo {author} {\bibfnamefont {T.}~\bibnamefont
  {Cao}}, \bibinfo {author} {\bibfnamefont {W.}~\bibnamefont {Bao}}, \bibinfo
  {author} {\bibfnamefont {C.}~\bibnamefont {Wang}}, \bibinfo {author}
  {\bibfnamefont {Y.}~\bibnamefont {Wang}}, \bibinfo {author} {\bibfnamefont
  {Z.~Q.}\ \bibnamefont {Qiu}}, \bibinfo {author} {\bibfnamefont {R.~J.}\
  \bibnamefont {Cava}}, \bibinfo {author} {\bibfnamefont {S.~G.}\ \bibnamefont
  {Louie}}, \bibinfo {author} {\bibfnamefont {J.}~\bibnamefont {Xia}}, \ and\
  \bibinfo {author} {\bibfnamefont {X.}~\bibnamefont {Zhang}},\ }\href@noop {}
  {\bibfield  {journal} {\bibinfo  {journal} {Nature}\ }\textbf {\bibinfo
  {volume} {546}},\ \bibinfo {pages} {265} (\bibinfo {year}
  {2017})}\BibitemShut {NoStop}%
\bibitem [{\citenamefont {Huang}\ \emph {et~al.}(2017)\citenamefont {Huang},
  \citenamefont {Clark}, \citenamefont {Navarro-Moratalla}, \citenamefont
  {Klein}, \citenamefont {Cheng}, \citenamefont {Seyler}, \citenamefont
  {Zhong}, \citenamefont {Schmidgall}, \citenamefont {McGuire}, \citenamefont
  {Cobden}, \citenamefont {Yao}, \citenamefont {Xiao}, \citenamefont
  {Jarillo-Herrero},\ and\ \citenamefont {Xu}}]{HuangXu2017}%
  \BibitemOpen
  \bibfield  {author} {\bibinfo {author} {\bibfnamefont {B.}~\bibnamefont
  {Huang}}, \bibinfo {author} {\bibfnamefont {G.}~\bibnamefont {Clark}},
  \bibinfo {author} {\bibfnamefont {E.}~\bibnamefont {Navarro-Moratalla}},
  \bibinfo {author} {\bibfnamefont {D.~R.}\ \bibnamefont {Klein}}, \bibinfo
  {author} {\bibfnamefont {R.}~\bibnamefont {Cheng}}, \bibinfo {author}
  {\bibfnamefont {K.~L.}\ \bibnamefont {Seyler}}, \bibinfo {author}
  {\bibfnamefont {D.}~\bibnamefont {Zhong}}, \bibinfo {author} {\bibfnamefont
  {E.}~\bibnamefont {Schmidgall}}, \bibinfo {author} {\bibfnamefont {M.~A.}\
  \bibnamefont {McGuire}}, \bibinfo {author} {\bibfnamefont {D.~H.}\
  \bibnamefont {Cobden}}, \bibinfo {author} {\bibfnamefont {W.}~\bibnamefont
  {Yao}}, \bibinfo {author} {\bibfnamefont {D.}~\bibnamefont {Xiao}}, \bibinfo
  {author} {\bibfnamefont {P.}~\bibnamefont {Jarillo-Herrero}}, \ and\ \bibinfo
  {author} {\bibfnamefont {X.}~\bibnamefont {Xu}},\ }\href@noop {} {\bibfield
  {journal} {\bibinfo  {journal} {Nature}\ }\textbf {\bibinfo {volume} {546}},\
  \bibinfo {pages} {270} (\bibinfo {year} {2017})}\BibitemShut {NoStop}%
\bibitem [{\citenamefont {Miller}(2017)}]{mag2d-Miller2017-pt}%
  \BibitemOpen
  \bibfield  {author} {\bibinfo {author} {\bibfnamefont {J.~L.}\ \bibnamefont
  {Miller}},\ }\href {\doibase 10.1063/PT.3.3613} {\bibfield  {journal}
  {\bibinfo  {journal} {Physics Today}\ }\textbf {\bibinfo {volume} {70}},\
  \bibinfo {pages} {16} (\bibinfo {year} {2017})}\BibitemShut {NoStop}%
\bibitem [{\citenamefont {Hao}\ \emph {et~al.}(2018)\citenamefont {Hao},
  \citenamefont {Meyers}, \citenamefont {Suwa}, \citenamefont {Yang},
  \citenamefont {Frederick}, \citenamefont {Dasa}, \citenamefont {Fabbris},
  \citenamefont {Horak}, \citenamefont {Kriegner}, \citenamefont {Choi},
  \citenamefont {Kim}, \citenamefont {Haskel}, \citenamefont {Ryan},
  \citenamefont {Xu}, \citenamefont {Batista}, \citenamefont {Dean},\ and\
  \citenamefont {Liu}}]{mag2d-Hao2018-nat}%
  \BibitemOpen
  \bibfield  {author} {\bibinfo {author} {\bibfnamefont {L.}~\bibnamefont
  {Hao}}, \bibinfo {author} {\bibfnamefont {D.}~\bibnamefont {Meyers}},
  \bibinfo {author} {\bibfnamefont {H.}~\bibnamefont {Suwa}}, \bibinfo {author}
  {\bibfnamefont {J.}~\bibnamefont {Yang}}, \bibinfo {author} {\bibfnamefont
  {C.}~\bibnamefont {Frederick}}, \bibinfo {author} {\bibfnamefont {T.~R.}\
  \bibnamefont {Dasa}}, \bibinfo {author} {\bibfnamefont {G.}~\bibnamefont
  {Fabbris}}, \bibinfo {author} {\bibfnamefont {L.}~\bibnamefont {Horak}},
  \bibinfo {author} {\bibfnamefont {D.}~\bibnamefont {Kriegner}}, \bibinfo
  {author} {\bibfnamefont {Y.}~\bibnamefont {Choi}}, \bibinfo {author}
  {\bibfnamefont {J.-W.}\ \bibnamefont {Kim}}, \bibinfo {author} {\bibfnamefont
  {D.}~\bibnamefont {Haskel}}, \bibinfo {author} {\bibfnamefont {P.~J.}\
  \bibnamefont {Ryan}}, \bibinfo {author} {\bibfnamefont {H.}~\bibnamefont
  {Xu}}, \bibinfo {author} {\bibfnamefont {C.~D.}\ \bibnamefont {Batista}},
  \bibinfo {author} {\bibfnamefont {M.~P.~M.}\ \bibnamefont {Dean}}, \ and\
  \bibinfo {author} {\bibfnamefont {J.}~\bibnamefont {Liu}},\ }\href {\doibase
  10.1038/s41567-018-0152-6} {\bibfield  {journal} {\bibinfo  {journal} {Nature
  Physics}\ }\textbf {\bibinfo {volume} {14}},\ \bibinfo {pages} {806}
  (\bibinfo {year} {2018})},\ \Eprint {http://arxiv.org/abs/1804.08780}
  {arXiv:1804.08780} \BibitemShut {NoStop}%
\bibitem [{\citenamefont {{Fei}}\ \emph {et~al.}(2018)\citenamefont {{Fei}},
  \citenamefont {{Huang}}, \citenamefont {{Malinowski}}, \citenamefont
  {{Wang}}, \citenamefont {{Song}}, \citenamefont {{Sanchez}}, \citenamefont
  {{Yao}}, \citenamefont {{Xiao}}, \citenamefont {{Zhu}}, \citenamefont
  {{May}}, \citenamefont {{Wu}}, \citenamefont {{Cobden}}, \citenamefont
  {{Chu}},\ and\ \citenamefont {{Xu}}}]{FeiXu2018}%
  \BibitemOpen
  \bibfield  {author} {\bibinfo {author} {\bibfnamefont {Z.}~\bibnamefont
  {{Fei}}}, \bibinfo {author} {\bibfnamefont {B.}~\bibnamefont {{Huang}}},
  \bibinfo {author} {\bibfnamefont {P.}~\bibnamefont {{Malinowski}}}, \bibinfo
  {author} {\bibfnamefont {W.}~\bibnamefont {{Wang}}}, \bibinfo {author}
  {\bibfnamefont {T.}~\bibnamefont {{Song}}}, \bibinfo {author} {\bibfnamefont
  {J.}~\bibnamefont {{Sanchez}}}, \bibinfo {author} {\bibfnamefont
  {W.}~\bibnamefont {{Yao}}}, \bibinfo {author} {\bibfnamefont
  {D.}~\bibnamefont {{Xiao}}}, \bibinfo {author} {\bibfnamefont
  {X.}~\bibnamefont {{Zhu}}}, \bibinfo {author} {\bibfnamefont
  {A.}~\bibnamefont {{May}}}, \bibinfo {author} {\bibfnamefont
  {W.}~\bibnamefont {{Wu}}}, \bibinfo {author} {\bibfnamefont {D.}~\bibnamefont
  {{Cobden}}}, \bibinfo {author} {\bibfnamefont {J.-H.}\ \bibnamefont {{Chu}}},
  \ and\ \bibinfo {author} {\bibfnamefont {X.}~\bibnamefont {{Xu}}},\
  }\href@noop {} {\bibfield  {journal} {\bibinfo  {journal} {Nature materials}\
  }\textbf {\bibinfo {volume} {17}},\ \bibinfo {pages} {778} (\bibinfo {year}
  {2018})}\BibitemShut {NoStop}%
\bibitem [{\citenamefont {Wang}\ \emph {et~al.}(2018)\citenamefont {Wang},
  \citenamefont {Guti{\'e}rrez-Lezama}, \citenamefont {Ubrig}, \citenamefont
  {Kroner}, \citenamefont {Gibertini}, \citenamefont {Taniguchi}, \citenamefont
  {Watanabe}, \citenamefont {Imamo{\u g}lu}, \citenamefont {Giannini},\ and\
  \citenamefont {Morpurgo}}]{WangMorpurgo2018}%
  \BibitemOpen
  \bibfield  {author} {\bibinfo {author} {\bibfnamefont {Z.}~\bibnamefont
  {Wang}}, \bibinfo {author} {\bibfnamefont {I.}~\bibnamefont
  {Guti{\'e}rrez-Lezama}}, \bibinfo {author} {\bibfnamefont {N.}~\bibnamefont
  {Ubrig}}, \bibinfo {author} {\bibfnamefont {M.}~\bibnamefont {Kroner}},
  \bibinfo {author} {\bibfnamefont {M.}~\bibnamefont {Gibertini}}, \bibinfo
  {author} {\bibfnamefont {T.}~\bibnamefont {Taniguchi}}, \bibinfo {author}
  {\bibfnamefont {K.}~\bibnamefont {Watanabe}}, \bibinfo {author}
  {\bibfnamefont {A.}~\bibnamefont {Imamo{\u g}lu}}, \bibinfo {author}
  {\bibfnamefont {E.}~\bibnamefont {Giannini}}, \ and\ \bibinfo {author}
  {\bibfnamefont {A.~F.}\ \bibnamefont {Morpurgo}},\ }\href@noop {} {\bibfield
  {journal} {\bibinfo  {journal} {Nature Communications}\ }\textbf {\bibinfo
  {volume} {9}},\ \bibinfo {pages} {2516} (\bibinfo {year} {2018})}\BibitemShut
  {NoStop}%
\bibitem [{\citenamefont {Huang}\ \emph {et~al.}(2018)\citenamefont {Huang},
  \citenamefont {Clark}, \citenamefont {Klein}, \citenamefont {MacNeill},
  \citenamefont {Navarro-Moratalla}, \citenamefont {Seyler}, \citenamefont
  {Wilson}, \citenamefont {McGuire}, \citenamefont {Cobden}, \citenamefont
  {Xiao}, \citenamefont {Yao}, \citenamefont {Jarillo-Herrero},\ and\
  \citenamefont {Xu}}]{HuangXu2018}%
  \BibitemOpen
  \bibfield  {author} {\bibinfo {author} {\bibfnamefont {B.}~\bibnamefont
  {Huang}}, \bibinfo {author} {\bibfnamefont {G.}~\bibnamefont {Clark}},
  \bibinfo {author} {\bibfnamefont {D.~R.}\ \bibnamefont {Klein}}, \bibinfo
  {author} {\bibfnamefont {D.}~\bibnamefont {MacNeill}}, \bibinfo {author}
  {\bibfnamefont {E.}~\bibnamefont {Navarro-Moratalla}}, \bibinfo {author}
  {\bibfnamefont {K.~L.}\ \bibnamefont {Seyler}}, \bibinfo {author}
  {\bibfnamefont {N.}~\bibnamefont {Wilson}}, \bibinfo {author} {\bibfnamefont
  {M.~A.}\ \bibnamefont {McGuire}}, \bibinfo {author} {\bibfnamefont {D.~H.}\
  \bibnamefont {Cobden}}, \bibinfo {author} {\bibfnamefont {D.}~\bibnamefont
  {Xiao}}, \bibinfo {author} {\bibfnamefont {W.}~\bibnamefont {Yao}}, \bibinfo
  {author} {\bibfnamefont {P.}~\bibnamefont {Jarillo-Herrero}}, \ and\ \bibinfo
  {author} {\bibfnamefont {X.}~\bibnamefont {Xu}},\ }\href@noop {} {\bibfield
  {journal} {\bibinfo  {journal} {Nature Nanotechnology}\ }\textbf {\bibinfo
  {volume} {13}},\ \bibinfo {pages} {544} (\bibinfo {year} {2018})}\BibitemShut
  {NoStop}%
\bibitem [{\citenamefont {Klein}\ \emph {et~al.}(2018)\citenamefont {Klein},
  \citenamefont {MacNeill}, \citenamefont {Lado}, \citenamefont {Soriano},
  \citenamefont {Navarro-Moratalla}, \citenamefont {Watanabe}, \citenamefont
  {Taniguchi}, \citenamefont {Manni}, \citenamefont {Canfield}, \citenamefont
  {Fern{\'a}ndez-Rossier},\ and\ \citenamefont
  {Jarillo-Herrero}}]{KleinJarillo2018}%
  \BibitemOpen
  \bibfield  {author} {\bibinfo {author} {\bibfnamefont {D.~R.}\ \bibnamefont
  {Klein}}, \bibinfo {author} {\bibfnamefont {D.}~\bibnamefont {MacNeill}},
  \bibinfo {author} {\bibfnamefont {J.~L.}\ \bibnamefont {Lado}}, \bibinfo
  {author} {\bibfnamefont {D.}~\bibnamefont {Soriano}}, \bibinfo {author}
  {\bibfnamefont {E.}~\bibnamefont {Navarro-Moratalla}}, \bibinfo {author}
  {\bibfnamefont {K.}~\bibnamefont {Watanabe}}, \bibinfo {author}
  {\bibfnamefont {T.}~\bibnamefont {Taniguchi}}, \bibinfo {author}
  {\bibfnamefont {S.}~\bibnamefont {Manni}}, \bibinfo {author} {\bibfnamefont
  {P.}~\bibnamefont {Canfield}}, \bibinfo {author} {\bibfnamefont
  {J.}~\bibnamefont {Fern{\'a}ndez-Rossier}}, \ and\ \bibinfo {author}
  {\bibfnamefont {P.}~\bibnamefont {Jarillo-Herrero}},\ }\href@noop {}
  {\bibfield  {journal} {\bibinfo  {journal} {Science}\ }\textbf {\bibinfo
  {volume} {360}},\ \bibinfo {pages} {1218} (\bibinfo {year}
  {2018})}\BibitemShut {NoStop}%
\bibitem [{\citenamefont {Song}\ \emph {et~al.}(2018)\citenamefont {Song},
  \citenamefont {Cai}, \citenamefont {Tu}, \citenamefont {Zhang}, \citenamefont
  {Huang}, \citenamefont {Wilson}, \citenamefont {Seyler}, \citenamefont {Zhu},
  \citenamefont {Taniguchi}, \citenamefont {Watanabe}, \citenamefont {McGuire},
  \citenamefont {Cobden}, \citenamefont {Xiao}, \citenamefont {Yao},\ and\
  \citenamefont {Xu}}]{SongXu2018}%
  \BibitemOpen
  \bibfield  {author} {\bibinfo {author} {\bibfnamefont {T.}~\bibnamefont
  {Song}}, \bibinfo {author} {\bibfnamefont {X.}~\bibnamefont {Cai}}, \bibinfo
  {author} {\bibfnamefont {M.~W.-Y.}\ \bibnamefont {Tu}}, \bibinfo {author}
  {\bibfnamefont {X.}~\bibnamefont {Zhang}}, \bibinfo {author} {\bibfnamefont
  {B.}~\bibnamefont {Huang}}, \bibinfo {author} {\bibfnamefont {N.~P.}\
  \bibnamefont {Wilson}}, \bibinfo {author} {\bibfnamefont {K.~L.}\
  \bibnamefont {Seyler}}, \bibinfo {author} {\bibfnamefont {L.}~\bibnamefont
  {Zhu}}, \bibinfo {author} {\bibfnamefont {T.}~\bibnamefont {Taniguchi}},
  \bibinfo {author} {\bibfnamefont {K.}~\bibnamefont {Watanabe}}, \bibinfo
  {author} {\bibfnamefont {M.~A.}\ \bibnamefont {McGuire}}, \bibinfo {author}
  {\bibfnamefont {D.~H.}\ \bibnamefont {Cobden}}, \bibinfo {author}
  {\bibfnamefont {D.}~\bibnamefont {Xiao}}, \bibinfo {author} {\bibfnamefont
  {W.}~\bibnamefont {Yao}}, \ and\ \bibinfo {author} {\bibfnamefont
  {X.}~\bibnamefont {Xu}},\ }\href@noop {} {\bibfield  {journal} {\bibinfo
  {journal} {Science}\ }\textbf {\bibinfo {volume} {360}},\ \bibinfo {pages}
  {1214} (\bibinfo {year} {2018})}\BibitemShut {NoStop}%
\bibitem [{\citenamefont {Jiang}\ \emph
  {et~al.}(2018{\natexlab{a}})\citenamefont {Jiang}, \citenamefont {Shan},\
  and\ \citenamefont {Mak}}]{JiangMak2018}%
  \BibitemOpen
  \bibfield  {author} {\bibinfo {author} {\bibfnamefont {S.}~\bibnamefont
  {Jiang}}, \bibinfo {author} {\bibfnamefont {J.}~\bibnamefont {Shan}}, \ and\
  \bibinfo {author} {\bibfnamefont {K.~F.}\ \bibnamefont {Mak}},\ }\href@noop
  {} {\bibfield  {journal} {\bibinfo  {journal} {Nature materials}\ }\textbf
  {\bibinfo {volume} {17}},\ \bibinfo {pages} {406} (\bibinfo {year}
  {2018}{\natexlab{a}})}\BibitemShut {NoStop}%
\bibitem [{\citenamefont {Jiang}\ \emph
  {et~al.}(2018{\natexlab{b}})\citenamefont {Jiang}, \citenamefont {Li},
  \citenamefont {Wang}, \citenamefont {Mak},\ and\ \citenamefont
  {Shan}}]{JiangShan2018}%
  \BibitemOpen
  \bibfield  {author} {\bibinfo {author} {\bibfnamefont {S.}~\bibnamefont
  {Jiang}}, \bibinfo {author} {\bibfnamefont {L.}~\bibnamefont {Li}}, \bibinfo
  {author} {\bibfnamefont {Z.}~\bibnamefont {Wang}}, \bibinfo {author}
  {\bibfnamefont {K.~F.}\ \bibnamefont {Mak}}, \ and\ \bibinfo {author}
  {\bibfnamefont {J.}~\bibnamefont {Shan}},\ }\href@noop {} {\bibfield
  {journal} {\bibinfo  {journal} {Nature Nanotechnology}\ }\textbf {\bibinfo
  {volume} {13}},\ \bibinfo {pages} {549} (\bibinfo {year}
  {2018}{\natexlab{b}})}\BibitemShut {NoStop}%
\bibitem [{\citenamefont {Seyler}\ \emph {et~al.}(2018)\citenamefont {Seyler},
  \citenamefont {Zhong}, \citenamefont {Klein}, \citenamefont {Gao},
  \citenamefont {Zhang}, \citenamefont {Huang}, \citenamefont
  {Navarro-Moratalla}, \citenamefont {Yang}, \citenamefont {Cobden},
  \citenamefont {McGuire}, \citenamefont {Yao}, \citenamefont {Xiao},
  \citenamefont {Jarillo-Herrero},\ and\ \citenamefont {Xu}}]{SeylerXu2018}%
  \BibitemOpen
  \bibfield  {author} {\bibinfo {author} {\bibfnamefont {K.~L.}\ \bibnamefont
  {Seyler}}, \bibinfo {author} {\bibfnamefont {D.}~\bibnamefont {Zhong}},
  \bibinfo {author} {\bibfnamefont {D.~R.}\ \bibnamefont {Klein}}, \bibinfo
  {author} {\bibfnamefont {S.}~\bibnamefont {Gao}}, \bibinfo {author}
  {\bibfnamefont {X.}~\bibnamefont {Zhang}}, \bibinfo {author} {\bibfnamefont
  {B.}~\bibnamefont {Huang}}, \bibinfo {author} {\bibfnamefont
  {E.}~\bibnamefont {Navarro-Moratalla}}, \bibinfo {author} {\bibfnamefont
  {L.}~\bibnamefont {Yang}}, \bibinfo {author} {\bibfnamefont {D.~H.}\
  \bibnamefont {Cobden}}, \bibinfo {author} {\bibfnamefont {M.~A.}\
  \bibnamefont {McGuire}}, \bibinfo {author} {\bibfnamefont {W.}~\bibnamefont
  {Yao}}, \bibinfo {author} {\bibfnamefont {D.}~\bibnamefont {Xiao}}, \bibinfo
  {author} {\bibfnamefont {P.}~\bibnamefont {Jarillo-Herrero}}, \ and\ \bibinfo
  {author} {\bibfnamefont {X.}~\bibnamefont {Xu}},\ }\href@noop {} {\bibfield
  {journal} {\bibinfo  {journal} {Nature Physics}\ }\textbf {\bibinfo {volume}
  {14}},\ \bibinfo {pages} {277} (\bibinfo {year} {2018})}\BibitemShut
  {NoStop}%
\end{thebibliography}
%

\end{document}